\documentclass[twocolumn]{aastex62}

\usepackage{mathtools}
\usepackage{upgreek}
\usepackage{dsfont}
\usepackage{natbib}

\defcitealias{bib:sidc}{SILSO WDC 2011-2018}

\newcommand{\tilt}{\ensuremath{\alpha}}
\newcommand{\cmf}{\ensuremath{B_{0}}}
\newcommand{\cmfi}[1]{\ensuremath{B_{0,#1}}}
\newcommand{\cpa}{\ensuremath{k_{\parallel}^{0}}}
\newcommand{\apar}{\ensuremath{a_{\parallel}}}
\newcommand{\bpar}{\ensuremath{b_{\parallel}}}
\newcommand{\aperp}{\ensuremath{a_{\perp}}}
\newcommand{\bperp}{\ensuremath{b_{\perp}}}

\newcommand{\fluxrig}{\ensuremath{/(\mathrm{m}^{2}\,\mathrm{sr}\,\mathrm{s}\,\mathrm{GV})}}
\newcommand{\GV}{\ensuremath{\mathrm{GV}}}

\newcommand{\diffcoeff}{\ensuremath{6\!\times\!10^{20}\ \mathrm{cm}^{2}/\mathrm{s}}}
\newcommand{\degree}{\ensuremath{^{\circ}}}

\newcommand{\ie}{\textit{i.e.}}
\newcommand{\eg}{\textit{e.g.}}

\newcommand{\Krr}{\ensuremath{\mathsf{K}_{rr}}}

\newcommand{\kpar}{\ensuremath{k_{\parallel}}}
\newcommand{\kperpr}{\ensuremath{k_{\perp,r}}}
\newcommand{\kperpt}{\ensuremath{k_{\perp,\theta}}}

\newcommand{\He}[1]{\ensuremath{{}^{#1}\mathrm{He}}}
\newcommand{\p}{\ensuremath{\mathrm{p}}}

\received{October 2018}
\revised{December 2018}
\accepted{???}
\submitjournal{ApJ}

\shorttitle{Numerical modeling of AMS-02 monthly p \& He}
\shortauthors{Corti et al.}

\begin{document}
   
\title{Numerical modeling of galactic cosmic ray proton and helium observed by AMS-02 during the solar maximum of Solar Cycle 24}

\correspondingauthor{Claudio Corti}
\email{corti@hawaii.edu}

\author[0000-0001-9127-7133]{Claudio Corti}
\affiliation{Physics and Astronomy Department, University of Hawaii at Manoa \\
   2505 Correa Road, \\
   Honolulu, HI 96822, USA}

\author[0000-0003-0793-7333]{Marius S. Potgieter}
\affiliation{Center for Space Research, North-West University, \\
   Potchefstroom, 2520, South-Africa}

\author[0000-0002-6706-0556]{Veronica Bindi}
\affiliation{Physics and Astronomy Department, University of Hawaii at Manoa \\
   2505 Correa Road, \\
   Honolulu, HI 96822, USA}

\author[0000-0003-4257-4187]{Cristina Consolandi}
\affiliation{Physics and Astronomy Department, University of Hawaii at Manoa \\
   2505 Correa Road, \\
   Honolulu, HI 96822, USA}

\author[0000-0002-8403-2004]{Christopher Light}
\affiliation{Physics and Astronomy Department, University of Hawaii at Manoa \\
   2505 Correa Road, \\
   Honolulu, HI 96822, USA}

\author[0000-0002-0135-8181]{Matteo Palermo}
\affiliation{Physics and Astronomy Department, University of Hawaii at Manoa \\
   2505 Correa Road, \\
   Honolulu, HI 96822, USA}

\author[0000-0002-3157-8839]{Alexis Popkow}
\affiliation{Physics and Astronomy Department, University of Hawaii at Manoa \\
   2505 Correa Road, \\
   Honolulu, HI 96822, USA}

\begin{abstract}
   Galactic cosmic rays (GCRs) are affected by solar modulation while they propagate through the heliosphere.
   The study of the time variation of GCR spectra observed at Earth can shed light on the underlying physical processes, specifically diffusion and particle drifts.
   Recently, the AMS-02 experiment measured with very high accuracy the time variation of the cosmic ray proton and helium flux between May 2011 and May 2017 in the rigidity range from 1 to 60 GV.
   In this work, a comprehensive three-dimensional (3D) steady-state numerical model is used to solve Parker's transport equation and is used to reproduce the monthly proton fluxes observed by AMS-02.
   We find that the rigidity slope of the perpendicular mean free path above 4 GV remains constant, while below 4 GV it increases during solar maximum.
   Assuming the same mean free paths for helium and protons, the models are able to reproduce the time behavior of the p/He ratio observed by AMS-02.
   The dependence of the diffusion tensor on the particle mass-to-charge ratio, $A/Z$, is found to be the main cause of the time dependence of p/He below 3 GV.
\end{abstract}

%
%
%
\section{Introduction} \label{sec:intro}
Galactic cosmic rays (GCRs) are charged particles produced by some of the most energetic phenomena in the Universe, which travel the endless voids of our galaxy before finally arriving at the edge of the solar system \citep{bib:amato17:gcr-review}.
Here they meet with the \emph{heliosphere}, a huge cavity carved out of the interstellar space by a supersonic stream of magnetized plasma constantly blown out from the Sun, called \emph{solar wind} \citep{bib:parker58:solar-wind}.
By the time the GCRs reach Earth, they have interacted with the turbulent magnetic field embedded in the time-varying solar wind: the overall effect of the physical processes involved in this interaction is called \emph{solar modulation} \citep{bib:parker65:modulation,bib:potgieter13:solar-modulation}.

In recent years, a new interest in GCRs has spurred from the observations of an excess in their anti-matter components, like positrons \citep{bib:adriani13:pamela-pos-flux,bib:aguilar14:ams-pos-frac} and anti-protons \citep{bib:adriani10:pamela-pbar-flux,bib:aguilar16:ams-pbarp-ratio}, suggesting an exotic origin, such as dark matter annihilation or decay \citep{bib:turner90:dm-pos,bib:donato09:dm-pbar}, or new astrophysical phenomena \citep{bib:hooper09:pwn-pos,bib:blum13:pos-secondary,bib:blasi09:sn-pbar}.
Since the fluxes of the various species of GCRs are distorted by the influence of the Sun below a few tens of GV, a better understanding of the solar modulation and its time evolution is of paramount importance to correctly deduce their shape before they enter the heliosphere \citep{bib:fornengo13:dbar-dm,bib:fornengo14:pbar-dm,bib:yuan15:pos-solmod,bib:cirelli14:pbar-dm,bib:tomassetti17:cr-prop-sol-nuc-unc}.

GCRs are also an unavoidable challenge for any human space exploration program, since they are a highly ionizing form of radiation, which can penetrate the walls of a spacecraft, an astronaut spacesuit and the human body itself \citep{bib:cucinotta06:space-radiation}.
The knowledge of the time variation of the GCR flux, and the study of the propagation of particles in the heliosphere, will help reduce the uncertainties in the radiation dose predictions \citep{bib:cucinotta13:radiation-risk}.

Recently, the AMS-02 experiment on board the International Space Station measured with very high accuracy the time variation, on a scale of a Bartels rotation (BR, 27 days), of the cosmic ray proton and helium flux between May 2011 and May 2017 in the rigidity range from 1 to 60 GV (\cite{bib:aguilar18:ams-monthly-phe}; data can also be retrieved at NASA's CDAWeb\footnote{\url{https://spdf.gsfc.nasa.gov/pub/data/international_space_station_iss/ams-02/}}).
The key points of AMS-02 observations are the complex time behavior due to the short-term activity and the decrease of the p/He ratio coinciding with the start of the flux recovery after the solar maximum.

In this work, we use a comprehensive three-dimensional (3D) numerical model to solve the propagation equation of GCRs in the heliosphere, in order to understand the physical processes underlying the AMS-02 results.
In the following sections the numerical model will be detailed, specifying the various ingredients needed to correctly describe the physics of the heliospheric transport of GCRs.
Then, the method to reproduce the proton monthly fluxes will be presented, together with the results.
Next, the p/He prediction from the best-fit models will be compared with data, and finally, we will perform a dedicated study to understand the origin of the p/He time dependence.
%
%
%
\section{Numerical model description} \label{sec:numerical-model}
A state-of-the-art 3D steady-state numerical model has been developed during the past years \citep{bib:potgieter14:pamela-modulation,bib:vos15:pamela-modulation} to solve the Parker equation of GCR transport in the heliosphere \citep{bib:parker65:modulation}:
\begin{equation} \label{eqn:numerical-model:parker}
\frac{\partial f}{\partial t} + \mathbf{V}_{sw} \cdot \boldsymbol{\nabla}f - \boldsymbol{\nabla} \cdot \left( \boldsymbol{\mathsf{K}} \boldsymbol{\nabla}f \right) - \frac{\boldsymbol{\nabla} \cdot \mathbf{V}_{sw}}{3}  R \frac{\partial f}{\partial R} = 0,
\end{equation}
where $f(\mathbf{r},R)$ is the omni-directional GCR distribution function, $\mathbf{V}_{sw}$ is the solar wind speed, $\boldsymbol{\mathsf{K}}$ is the diffusion tensor, which can be separated into a symmetric part, describing the scattering of particles on the heliospheric magnetic field (HMF) irregularities, and an asymmetric part, describing particle drifts along magnetic field gradients, curvatures, and the heliospheric current sheet (HCS).
In the steady-state approximation, $\partial f/\partial t = 0$: this is a reasonable assumption during the solar minimum, but less so during the solar maximum.
Nevertheless, for studies of time variation of GCR fluxes averaged over BRs it is still acceptable.

The model uses a finite-difference solver, the alternating direction implicit (ADI) method \citep{bib:peaceman55:2d-adi}, to obtain $f$ at all positions in the heliosphere.
This method has been adapted to cope with four numerical dimensions: three spatial (therefore called 3D) and one to handle rigidity.
Including also a time-dependence would make the method numerically unsuitable so that either one spatial dimension should be sacrificed \citep{bib:ngobeni14:sol-mod-c} or the so-called SDE (stochastic differential equation) approach should be followed (see \eg\ \cite{bib:kopp17:transp-mag-structs}, \cite{bib:luo17:fd-sde}, and references therein).
%
%
%
\subsection{Solar wind, heliospheric magnetic field and current sheet}\label{sec:numerical-model:sw-hmf-hcs}
The solar wind velocity profile is assumed to be separable in a radial and latitudinal component:
\begin{equation} \label{eqn:numerical-model:vsw}
\mathbf{V}_{sw}(r,\,\theta) = V_{r}(r)V_{\theta}(\theta)\hat{\mathbf{r}}.
\end{equation}
The radial component describes the fast rise to supersonic speed within the first 0.3 AU from the Sun (first term of Equation \ref{eqn:numerical-model:vsw-rad}) and the transition to subsonic speed at the termination shock (second term of Equation \ref{eqn:numerical-model:vsw-rad}):
\begin{equation} \label{eqn:numerical-model:vsw-rad}
\begin{split}
V_{r}(r) & = 1 - \exp\left[ \frac{40}{3} \left( \frac{r_{\odot} - r}{r_{0}} \right) \right] \\
& + \left[ \frac{s + 1}{2 s} - \frac{s - 1}{2 s} \tanh \left( \frac{r - r_{TS}}{L} \right) - 1 \right],
\end{split}
\end{equation}
where $r_{\odot} = 0.005$ AU is the Sun radius, $r_{0} = 1$ AU, $r_{TS}$ is the radial position of the termination shock (which, in principle, can vary in time), $L=1.2$ AU is the width of the shock barrier and $s = 2.5$ is the shock compression ratio in the downstream region, \ie\ the ratio of the velocity before and after the shock.\\
The latitudinal term describes the transition between the slow (polar) and fast (equatorial) component of the solar wind:
\begin{equation} \label{eqn:numerical-model:vsw-lat}
V_{\theta}(\theta) = \frac{V_{pol} + V_{eq}}{2} \mp \frac{V_{pol} - V_{eq}}{2} \tanh \left[ 6.8 \left( \theta' \pm \xi \right)\right],
\end{equation}
where $V_{pol}$ and $V_{eq}$ are, respectively, the polar and equatorial solar wind speed components, $\theta' = \theta - \pi/2$ and $\xi$ is the polar angle at which the transition between the equatorial and polar streams begins.
The top and bottom signs correspond, respectively, to the northern ($0 < \theta < \pi/2$) and southern ($\pi/2 < \theta < \pi$) hemisphere.
During periods of solar maximum, there is no clear latitudinal dependence of the solar wind speed, so that on average $V_{pol} = V_{eq}$ and the second term of Equation \ref{eqn:numerical-model:vsw-lat} vanishes.\\
The reacceleration at the termination shock via diffusive shock acceleration (DSA) is not included in the model, since for protons above 1 GV the effects of the termination shock at Earth are negligible (see \eg\ \cite{bib:langner05:pro-ts-effect}, and references therein).
The drop in solar wind velocity at the termination shock is taken into account in the evaluation of the HMF and the diffusion tensor, reproducing the actual diffusion barrier present at the shock.

The HMF implemented in this model is the Parker field with the Smith-Bieber modification:
\begin{equation} \label{eqn:numerical-model:hmf}
\begin{split}
\mathbf{B}(r,\,\theta,\,\phi) & = B_{n} \left( \frac{r}{r_{0}} \right)^{2} \left( \hat{\mathbf{r}} - \tan\psi \hat{\boldsymbol{\upphi}} \right) \\
& \times \left[ 1 - 2H(\theta - \theta_{HCS})\right],
\end{split}
\end{equation}
where $B_{n}$ is a normalization factor dependent on the observed magnitude of the HMF at Earth, $B_{0}$; $H$ is the Heaviside step function, which describes the opposite polarity above and below the HCS; $\theta_{HCS}$ is the polar position of the HCS; $\psi$ is the spiral angle, \ie\ the angle between the direction of the HMF and the radial direction.
$\psi$ is defined as:
\begin{equation} \label{eqn:numerical-model:tanpsi}
\tan\psi = \frac{\Omega (r - b) \sin\theta}{V_{sw}(r,\,\theta)} - \frac{r}{b} \frac{V_{sw}(b,\,\theta)}{V_{sw}(r,\,\theta)} \frac{B_{T}(b)}{B_{R}(b)},
\end{equation}
where $\Omega$ is the angular rotation frequency of the Sun, $b = 20~r_{\odot}$ is the distance from the Sun where the HMF becomes fully radial and $B_{T}(b)/B_{R}(b) \approx -0.02$ is the ratio of the azimuthal-to-radial magnetic field components \citep{bib:smith91:hmf-sb-mod}.
Imposing $B(r_{0},\,\pi/2) = B_{0}$, we obtain $B_{n} = B_{0} / \sqrt{1 + \tan\psi(r_{0},\,\pi/2)}$.
See also \cite{bib:raath16:hmf-modifications} for a detailed study of the Smith-Bieber and other HMF modifications.

The position of the HCS is given by \cite{bib:kota83:drift-3d-model}:
\begin{equation} \label{eqn:numerical-model:theta-hcs}
\theta_{HCS} = \frac{\pi}{2} - \tan^{-1} \left[ \tan\alpha \sin \left( \phi + \Omega\frac{r - r_{\odot}}{V_{sw}(r,\,\theta)} \right) \right],
\end{equation}
where $\alpha$ is the tilt angle, \ie\ the maximum latitudinal extent of the HCS.
To avoid numerical instabilities created by the discontinuity of the polarity flip when passing from one side of the HCS to the other, the Heaviside function is replaced with a smooth transition function:
\begin{equation} \label{eqn:numerical-model:hcs}
A \tanh \left( 0.549 \frac{\theta_{HCS} - \theta}{\Delta\theta_{HCS}} \right),
\end{equation}
where $A$ is the HMF polarity ($\pm 1$) and $\Delta\theta_{HCS} = 2 r_{L}/r = 2 R/(r B c)$ is the angle spanned by two gyroradii for a particle with rigidity $R$.
This means that the HCS drift effects are taken into account only if the particle is within 2 gyroradii from the HCS.
See also \cite{bib:raath15:hcs} for a detailed study of how the treatment of the HCS in numerical modeling studies affects cosmic ray modulation.
%
%
%
\subsection{Diffusion and drift coefficients} \label{sec:numerical-model:diff-drift-coeff}
The rigidity dependence of the parallel diffusion coefficient is approximated by a double power-law with a smooth change of slope, while the radial dependence is assumed to be inversely proportional to the magnitude of the HMF:
\begin{equation} \label{eqn:numerical-model:kpar}
k_{\parallel} = \cpa \beta \frac{1\,\mathrm{nT}}{B} \left( \frac{R}{R_{k}} \right)^{a} \left[ 1 + \left( \frac{R}{R_{k}} \right)^{s} \right]^{\frac{b-a}{s}},
\end{equation}
where $\cpa$ is a normalization factor, $\beta = v/c$, $R_{k}$ is the rigidity at which the transition between the two power-laws happens, $a$ and $b$ are, respectively, the slopes of the low- and high-rigidity power-laws and $s$ controls the smoothness of the transition.
The perpendicular diffusion coefficients are assumed to be proportional to the parallel diffusion coefficient:
\begin{equation} \label{eqn:numerical-model:kperp}
k_{\perp,r} = k_{\perp,r}^{0} k_{\parallel} \quad \quad k_{\perp,\theta} = u(\theta) k_{\perp,\theta}^{0} k_{\parallel}.
\end{equation}
where $k_{\perp,r}^{0}$ and $k_{\perp,\theta}^{0}$ are scaling factors of the order of percent, while $u(\theta)$ is a function that enhances the perpendicular diffusion in the polar regions and it's defined as:
\begin{equation} \label{eqn:numerical-model:kperp-pol}
u(\theta) = \frac{3}{2} + \frac{1}{2} \tanh \left[ 8 \left( \left| \theta - \frac{\pi}{2} \right| - 35 \degree \right) \right].
\end{equation}
The numerical values are chosen to reproduce cosmic ray observations at higher latitudes by the \emph{Ulysses} spacecraft \citep{bib:potgieter93:ulysses,bib:kota95:anisotropy,bib:potgieter00:perp-enhancement,bib:heber06:modulation,bib:potgieter13:insights}.
We note that forcing on \kperpr\ and \kperpt\ the same rigidity dependence of \kpar\ is a simplification, since both turbulence theory and observations predict a different rigidity behavior (see \eg\ \cite{bib:burger00:lat-gradients}, and references therein).
In this work, the slopes of the perpendicular diffusion coefficient are not constrained to be equal to the ones of the parallel diffusion coefficient, therefore we introduce the parameters \apar, \bpar\ (slopes of the parallel diffusion coefficient) and \aperp, \bperp\ (slopes of the perpendicular diffusion coefficient).
The transition rigidity $R_{k}$ and the smoothness factor $s$ are assumed instead to be the same for all diffusion coefficients; for an overview of these aspects, see \cite{bib:potgieter17:global-mod}.

The drift coefficient is defined as:
\begin{equation} \label{eqn:numerical-model:drift}
k_{A} = k_{A}^{0} \frac{\beta R}{3 B} \frac{\left( R/R_{A} \right)^{2}}{1 + \left( R/R_{A} \right)^{2}},
\end{equation}
where $k_{A}^{0}$ is a normalization factor that can be used to reduce the overall drift effects, while $R_{A}$ is the rigidity below which the drift is suppressed due to scattering.
For a detailed study of how this expression is obtained and what effects it has on solar modulation of GCRs, see \cite{bib:ngobeni15:drift-scattering} and \cite{bib:nndanganeni16:drift-electrons}.
This approach means that the model is \emph{diffusion dominated}, rather than \emph{drift dominated} as the original drift models of the 1980s and 1990s were, and also as recently applied by \eg\ \cite{bib:tomassetti17:time-lag}.
%
%
%
\subsection{Local interstellar spectrum} \label{sec:numerical-model:LIS}
The proton and helium local interstellar spectrum (LIS) are parametrized between 0.1 GV and 3 TV as a combination of four smooth power-laws in rigidity:
\begin{equation} \label{eqn:lis}
\dfrac{dJ_{LIS}}{dR} = N \left( \dfrac{R}{1\,\mathrm{GV}} \right)^{\gamma_{0}} \prod_{i=1}^{3} \left[ \dfrac{1 + (R/R_{i})^{s_{i}}}{1 + R_{i}^{-s_{i}}} \right]^{\Delta_{i}/s_{i}},
\end{equation}
where $N$ is the flux normalization at 1 GV, $\gamma_{0}$ is the spectral index of the first power-law, $\Delta_{i} = \gamma_{i}-\gamma_{i-1}$ is the difference in spectral index between the $i$-th power-law and the previous one, $R_{i}$ are the rigidities at which the breaks between power-laws happen, and $s_{i}$ control the smoothness of the breaks.

Following the same method as in \cite{bib:corti16:lis-modffa}, the proton LIS is derived by a combined fit on low-rigidity data measured by Voyager 1 outside the heliosphere \citep{bib:stone13:voyager1-ism-cr} and high-rigidity data measured by AMS-02 \citep{bib:aguilar15:ams-p-flux}, modulated with the force-field approximation \citep{bib:gleeson68:ffa}.
The best-fit parameters are: $N = (5658 \pm 57)\fluxrig$, $\gamma_{0} = 1.669 \pm 0.005$, $R_{1} = (0.572 \pm 0.004)\ \GV$, $\Delta_{1} = -4.117 \pm 0.005$, $s_{1} = 1.78 \pm 0.02$, $R_{2} = (6.2 \pm 0.2)\ \GV$, $\Delta_{2} = -0.423 \pm 0.008$, $s_{2} = 3.89 \pm 0.49$, $R_{3} = (540 \pm 240)\ \GV$, $\Delta_{3} = -0.26 \pm 0.1$, and $s_{3} = 1.53 \pm 0.43$.
This LIS is consistent within $0.2\%$ with the one from \cite{bib:corti16:lis-modffa}.

\He3\ and \He4\ LIS are derived by a combined fit to multiple datasets\footnote{IMAX, BESS, AMS-01 and PAMELA data have been downloaded from the CRDB \citep{bib:maurin14:crdb}: \url{https://lpsc.in2p3.fr/cosmic-rays-db}.}: Voyager 1 He \citep{bib:cummings16:voyager1-ism-cr}, Voyager 1 \He3\ and \He4 \citep{bib:webber18:voyager1-ism-iso}, IMAX \He3/\He4 \citep{bib:reimer98:imax-he-iso}, BESS \He3\ and \He4 \citep{bib:wang02:bess-hheiso-flux,bib:myers03:bess-he-iso}, AMS-01 \He3\ and \He4 \citep{bib:aguilar11:ams01-iso}, PAMELA \He3\ and \He4 \citep{bib:adriani16:pamela-iso}, and AMS-02 He \citep{bib:aguilar17:ams-heco}.
Voyager 1 data were measured outside the heliosphere, while all other data were collected at 1 AU at different solar activity conditions, so they were modulated with the force-field approximation.
We allowed the modulation parameter for \He3\ to be different from the modulation parameter for \He4 (see Section \ref{sec:phe-comparison} for the dependence of the results on this assumption).
According to the standard model of GCR production, acceleration and transport in the galaxy, \He4 is produced in astrophysical sources, while \He3\ is produced by collisions of heavier nuclei with the interstellar material, so that \He3/\He4\ at very high-rigidity ($\gtrsim 100$ GV) becomes proportional to $1/D$, where $D \propto R^{\delta}$ is the diffusion coefficient in the galaxy \citep{bib:amato17:gcr-review}.
The latest B/C data from AMS-02 \citep{bib:aguilar16:ams-bc-ratio} constrain $\delta$ to be $-1/3$, in agreement with Kolmogorov theory of interstellar turbulence \citep{bib:kolmogorov41:turbulence}.
\begin{figure}[b]
   \centering
   \includegraphics[width=\columnwidth]{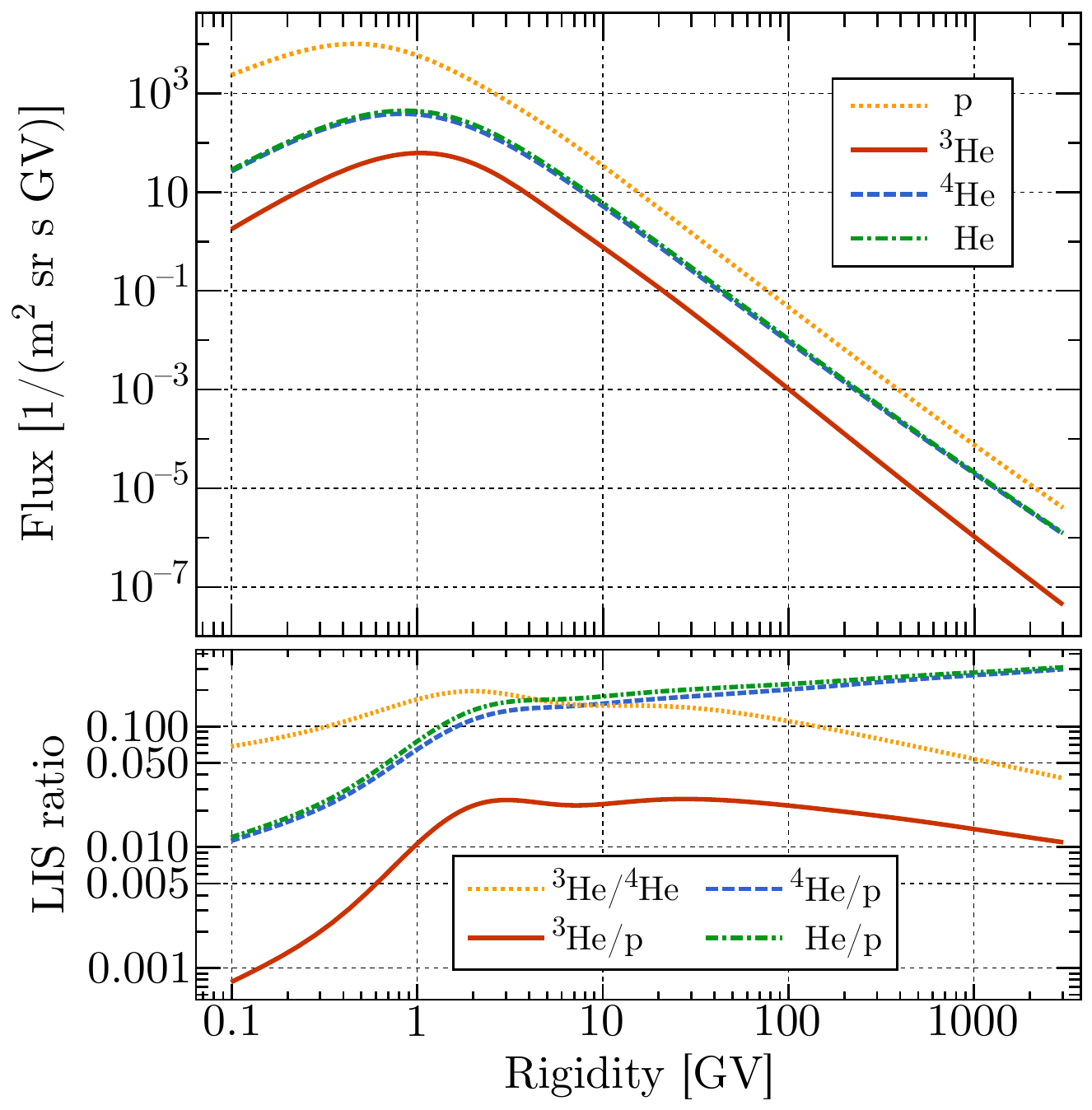}
   \caption
   {
      \textbf{Top:} p (dotted yellow), \He3 (solid red), \He4 (dashed blue), and He (dotted-dashed green) LIS parametrizations used in this paper, derived by a combined fit to Voyager 1 unmodulated data and various modulated datasets collected at 1 AU at different times (see text for details).
      \textbf{Bottom:} \He3/\He4 (dotted yellow), \He3/p (solid red), \He4/p (dashed blue), and He/p (dotted-dashed green) LIS ratio.
   }
   \label{fig:numerical-model:LIS}
\end{figure}
Furthermore, at high rigidities propagation in the galaxy should be mostly dependent on rigidity only, while at low rigidity energy loss processes are also velocity dependent.
For these reasons, the parameters $R_{2}$, $s_{2}$, $R_{3}$, $\Delta_{3}$, and $s_{3}$ for \He3\ are assumed to be equal to the ones for \He4, while $\gamma_{2}(\He3) = \gamma_{2}(\He4)-1/3$.
The parameters $N$, $\gamma_{0}$, $R_{1}$, $\Delta_{1}$, and $s_{1}$ are instead left free independently for \He3\ and \He4.
The best-fit parameters for \He4\ are: $N = (362 \pm 4)\fluxrig$, $\gamma_{0} = 2.113 \pm 0.007$, $R_{1} = (1.15 \pm 0.01)\ \GV$, $\Delta_{1} = -5.79 \pm 0.01$, $s_{1} = 1.27 \pm 0.01$, $R_{2} = (5.2 \pm 0.5)\ \GV$, $\Delta_{2} = 0.47 \pm 0.01$, $s_{2} = 2.19 \pm 0.06$, $R_{3} = (298 \pm 38)\ \GV$, $\Delta_{3} = 1.063 \pm 0.003$, and $s_{3} = 0.270 \pm 0.008$.
The best-fit parameters for \He3\ are: $N = (60.2 \pm 1.5)\fluxrig$, $\gamma_{0} = 2.29 \pm 0.04$, $R_{1} = (2.37 \pm 0.08)\ \GV$, $\Delta_{1} = -10 \pm 0.9$, $s_{1} = 1.27 \pm 0.06$.

Figure \ref{fig:numerical-model:LIS} shows a comparison of the p, \He3, \He4, and He (equal to \He3+\He4) \ LIS parametrizations (top panel) and their ratios (bottom panel).
For alternative methods of obtaining the proton and helium LIS, see \cite{bib:bisschoff16:lis}, and for a discussion of the impact of the Voyager and PAMELA observations on determining the appropriate LIS, see \cite{bib:potgieter14:lis}.
%
%
%
\section{Reproduction and fit of the AMS-02 monthly proton fluxes} \label{sec:fitting-strategy}
The standard approach of a least-squares fit with \texttt{MINUIT} \citep{bib:james75:minuit} is not feasible in this work, since a single model runs too slowly to allow the thousands of sequential iterations needed to find a global minimum.
Furthermore, the fit should be repeated for each of the 79 BRs observed by AMS-02, potentially generating a given model multiple times, thus wasting computing time.
To solve this issue, a different strategy has been developed.\\
An ensemble of models is created in parallel, each with a different combination of input parameters.
The resulting multi-dimensional grid of models is linearly interpolated to find the set of parameters that minimizes the chi-squared between the models and the data.
This way, the models are generated only once, and they can be reused in the fitting of every flux, avoiding their duplication.
The parameters and their values defining the multi-dimensional grid are listed in Table \ref{tab:fitting-strategy:parameters-grid}.

\begin{deluxetable}{m{12em}Cm{8em}}[t]
   \tablecaption{
      Definition of the grid of input parameters used to generate the numerical models. \label{tab:fitting-strategy:parameters-grid}
   }
   \tablehead{
      \colhead{Parameter} & \colhead{Symbol} & \colhead{Values}
   }
   \startdata
   HMF polarity                             & A       & $<0$, $>0$ \\
   Tilt angle (degrees)                     & \tilt   & 20, 25, 30, 35, 40, 55, 65, 75 \\
   HMF magnitude at Earth (nT)              & \cmf    & 4.5, 5.5, 6.0, 6.5, 7.5, 8.5 \\
   Normalization of the parallel 
   diffusion coefficient (\diffcoeff)       & \cpa    & 50, 70, 90, 110, 130, 150, 170, 190, 210, 230\tablenotemark{a}, 250\tablenotemark{a} \\
   Low-rigidity slope of the parallel
   diffusion coefficient                    & \apar   & 0.2, 0.5, 0.8, 1.1, 1.4, 1.7, 2.0 \\
   High-rigidity slope of the parallel
   diffusion coefficient                    & \bpar   & 0.2, 0.5, 0.8, 1.1, 1.4, 1.7, 2.0, 2.3 \\
   Low-rigidity slope of the perpendicular
   diffusion coefficient                    & \aperp  & 0.2, 0.5, 0.8, 1.1, 1.4, 1.7, 2.0 \\
   High-rigidity slope of the perpendicular
   diffusion coefficient                    & \bperp  & 0.2, 0.5, 0.8, 1.1, 1.4, 1.7, 2.0, 2.3 \\
   \enddata
   \tablenotetext{a}{Only for $A>0$ models.}
\end{deluxetable}
The normalization of the perpendicular radial and polar diffusion coefficients has been kept fixed at $k_{\perp,r}^{0} = 0.02$ and $k_{\perp,\theta}^{0} = 0.01$, consistent with the values found by \cite{bib:vos15:pamela-modulation}, \cite{bib:zhao14:mod-minimum-cycle23-24}, and \cite{bib:potgieter17:diff-mod-pro-ele} analyzing data from PAMELA and with the expectation of turbulence theory (see \eg\ \cite{bib:burger00:lat-gradients} and \cite{bib:bieber04:nlgct}).
The parameters describing the drift processes, $R_{A}$ and $k_{A}^{0}$, are set to the values used for reproducing PAMELA data, \ie\ 0.55 GV and 1, respectively.
The transition rigidity $R_{k}$ and the smoothness of the change of slope $s$ are the same for all the three diffusion coefficients and are equal to 4.3 GV and 2.2, respectively.
The termination shock is fixed at 80 AU and the heliopause at 122 AU, consistent with the \emph{Voyager} observations.
The equatorial and fast solar wind components have been assumed to have the same speed $V_{0} = 440$ km/s, since we are analyzing mostly the solar maximum period.\\
The spatial grid has 609 steps in the radial direction, from 0.4 AU to 122 AU, 145 steps in the polar direction, from 0 to $\pi$, and 33 steps in the longitudinal direction, from 0 to $2\pi$.
The rigidity grid has been divided in 245 steps, uniformly distributed in logarithmic space between 1 and 200 GV.
To reduce the output file size, the solution has been saved in a reduced spatial grid, with a radial step of 2 AU, a latitudinal step of 5 \degree and at $\phi=0$.
The latter choice is justified by the fact that the modulated flux at Earth is negligibly dependent on the heliographic longitude: indeed, the flux variation around the average value is of the order of 0.3\%.\\
More than 3 million models have been generated, for a running time of ten weeks and a total disk size of 4.6 TB.
%
%
%
\subsection{Heliosphere status} \label{sec:fitting-strategy:heliosphere-status}
A steady-state model assumes that the heliosphere status is frozen in the whole time interval during which the particles propagate from the heliopause to Earth.
Clearly, this assumption is never valid in a dynamical system like the heliosphere, especially during periods of high solar activity, when the HMF and the tilt angle can have large variations on a monthly basis.
Nevertheless, the steady-state approximation is widely used, due to the simplicity of treatment of the numerical solution of the Parker equation (see \eg\ \cite{bib:potgieter14:pamela-modulation}, \cite{bib:zhao14:mod-minimum-cycle23-24}, and \cite{bib:vos15:pamela-modulation}).\\
As a first approximation, a way to take into account the time-varying status of the heliosphere is to use an average value for \tilt\ and \cmf.
Given a BR, we take the average of the tilt angle and HMF over a time period preceding the selected BR.
This time period has been chosen such that the average values of \tilt\ and \cmf\ reflect the average conditions sampled by GCRs while propagating from the heliopause to Earth.
Since the HMF is frozen in the solar wind, it propagates with the same velocity: if $V_{0} = 440$ km/s, taking into account the drop in velocity at the termination shock, the propagation time is of the order of two years.
However, GCRs diffuse inward in a much shorter period of time, between 1 and 4 months \citep{bib:strauss11:prop-time-ene-loss}, and do not spend the same amount of time at all radial distances.
In fact, the more they penetrate the heliosphere, the more energy they lose, so that the residing time increases going toward the Sun.
At the same time, most of the modulation happens in the heliosheath, as observed by \emph{Voyager 1} (see \eg\ \cite{bib:webber12:voyager1-increase} and \cite{bib:vos15:pamela-modulation}).
We decided to consider a period of one year, during which the heliosphere conditions affect the GCRs.
See Section \ref{sec:results} for a discussion of the dependence of the results on the duration of this period.

\begin{figure}[t]
   \centering
   \includegraphics[width=\columnwidth]{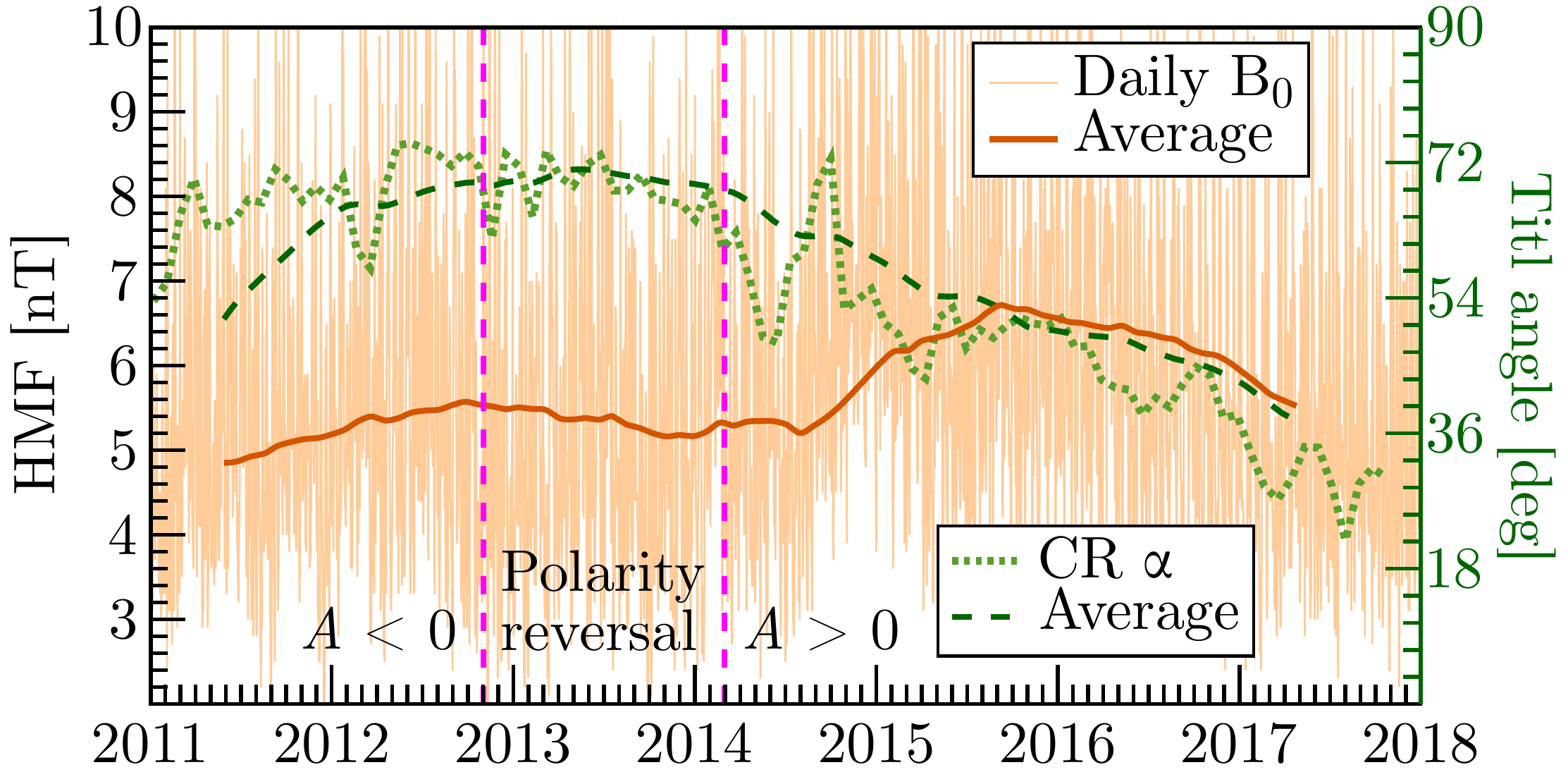}
   \caption
   {
      Time variation of the tilt angle, $\tilt$, measured in Carrington rotations by the WSO (light green dotted line) and of the daily HMF magnitude, $\cmf$, obtained by OMNIWeb (thin orange line).
      The thick dashed dark green and thick brown lines are the 1-year backward average of, respectively, the tilt angle and the HMF for every BR.
      The vertical dashed magenta lines delimit the period of the solar magnetic field polarity reversal.
   }
   \label{fig:fitting-strategy:heliosphere-status}
\end{figure}

Figure \ref{fig:fitting-strategy:heliosphere-status} illustrates the time variation of the tilt angle, measured every Carrington rotation by the Wilcox Solar Observatory\footnote{We used the classic model (line-of-sight) from \url{http://wso.stanford.edu/Tilts.html} \citep{bib:hoeksema95:tilt-angle}.} (WSO), and of the daily HMF observed at 1 AU by the ACE and \emph{Wind} spacecraft \footnote{The HMF magnitude data have been downloaded by the NASA/GSFC's OMNI data set through OMNIWeb: \url{https://omniweb.gsfc.nasa.gov/index.html}.}.
The values used as input for the models are computed with a 1-year backward average and are shown in thick lines.

For each BR, the tilt angle and the HMF are fixed to the values $\langle\tilt\rangle$ and $\langle\cmf\rangle$ obtained by the 1-year backward average.
Since the grid has only a few discrete values of \tilt\ and \cmf, a two-dimensional linear interpolation is used to obtain the modulated flux $\Phi(\langle\tilt\rangle,\,\langle\cmf\rangle)$ corresponding to the averaged heliosphere status.
Let's define $\Phi_{i,j} = \Phi(\,\tilt_{i},\,\cmfi{j};\mathbf{Q})$, where $\mathbf{Q} = (\cpa,\, \apar,\, \bpar,\, \aperp,\, \bperp)$ is a vector representing one of the possible combinations of the remaining parameters of the grid, while $i$ and $j$ are the points on, respectively, the \tilt\ axis and the \cmf\ axis for which $\tilt_{i} \leq \langle\tilt\rangle \leq \tilt_{i+1}$ and $\cmfi{j} \leq \langle\cmf\rangle \leq \cmfi{j+1}$.
Let's also define the interpolating factors $s_{\tilt} = (\langle\tilt\rangle - \tilt_{i})/(\tilt_{i+1} - \tilt_{i})$ and $s_{\cmf} = (\langle\cmf\rangle - \cmfi{i})/(\cmfi{i+1} - \cmfi{i})$.
If $\langle\tilt\rangle$ or $\langle\cmf\rangle$ is outside the range covered by the generated grid, then $s_{\tilt}$ and $s_{\cmf}$ are computed using the two closest points to $\langle\tilt\rangle$ and $\langle\cmf\rangle$.
A first interpolation is performed on the \cmf\ axis: $\Phi_{i}(\langle\cmf\rangle) = (1-s_{\cmf})\Phi_{i,j} + s_{\cmf}\Phi_{i,j+1}$ and $\Phi_{i+1}(\cmf) = (1-s_{\cmf})\Phi_{i+1,j} + s_{\cmf}\Phi_{i+1,j+1}$.
The final interpolation is carried out on the \tilt\ axis: $\Phi(\langle\tilt\rangle,\,\langle\cmf\rangle) = (1-s_{\tilt})\Phi_{i}(\langle\cmf\rangle) + s_{\tilt}\Phi_{i+1}(\langle\cmf\rangle)$.
The procedure is repeated for all the grid combinations of $\mathbf{Q}$.
We verified in a small sub-sample of the models that the 2D linear interpolation does not introduce any bias in the fluxes with respect to generating a model directly with $\langle\tilt\rangle$ and $\langle\cmf\rangle$: the difference due to the interpolation procedure is always much smaller than 1\%.

A good fraction of the monthly fluxes have been collected by AMS-02 during the period of the magnetic field polarity reversal.
Since the model expects a well-defined polarity, it's not possible to correctly describe the heliosphere status in this time interval.
For this reason, both models with negative and positive polarity have been used to describe the BRs between October 2013 and February 2015, while before October 2013 the polarity was only negative and after February 2015 only positive.
The reversal period ended in February 2104, but we decided to extend it up to one year later to take into account the propagation through the heliosphere.
%
%
%
\subsection{Best-fit parameters estimation} \label{sec:fitting-strategy:parameter-estimation}
The interpolated fluxes are used to estimate the best-fit parameters \cpa, \apar, \bpar, \aperp, and \bperp, and their time variation.
For every BR $n$ and for every model $m$ (with the corresponding set of parameters $\mathbf{Q}_{m}$), the chi-squared $\chi^{2}_{n,m}$ between the generated flux $\Phi_{n,m} = \Phi(\langle\tilt\rangle_{n},\, \langle\cmf\rangle_{n};\, \mathbf{Q}_{m})$ and the flux $F_{n}$ measured by AMS-02 is computed:
\begin{equation}
\chi^{2}_{n,m} = \sum_{i} \left( \frac{F_{n,i} - \Phi_{n,m}(\widetilde{R}_{i})}{\sigma_{n,i}} \right)^{2},
\end{equation}
where $i$ is the rigidity binning index of the AMS-02 data, and $\sigma_{n,i}$ is the AMS-02 uncertainty in the $i$-th rigidity bin.
The generated flux $\Phi_{n,m}$ is evaluated at the rigidity $\widetilde{R}_{i} = \sqrt{R_{i}R_{i+1}}$, where $R_{i}$ and $R_{i+1}$ are the left and right edge of the $i$-th rigidity bin, by interpolating the flux value between consecutive rigidity steps with a power-law.
The model $\widehat{m}(n)$ with the minimum chi-squared, $\widehat{\chi}^{2}_{n} = \chi^{2}_{n,\widehat{m}(n)}$, is considered as the best-fit model for the $n$-th BR, and the corresponding parameters, $\widehat{\mathbf{Q}}_{n} = \mathbf{Q}_{\widehat{m}(n)}$, as the best-fit parameters.

\begin{figure}[!h]
   \centering
   \includegraphics[width=\columnwidth]{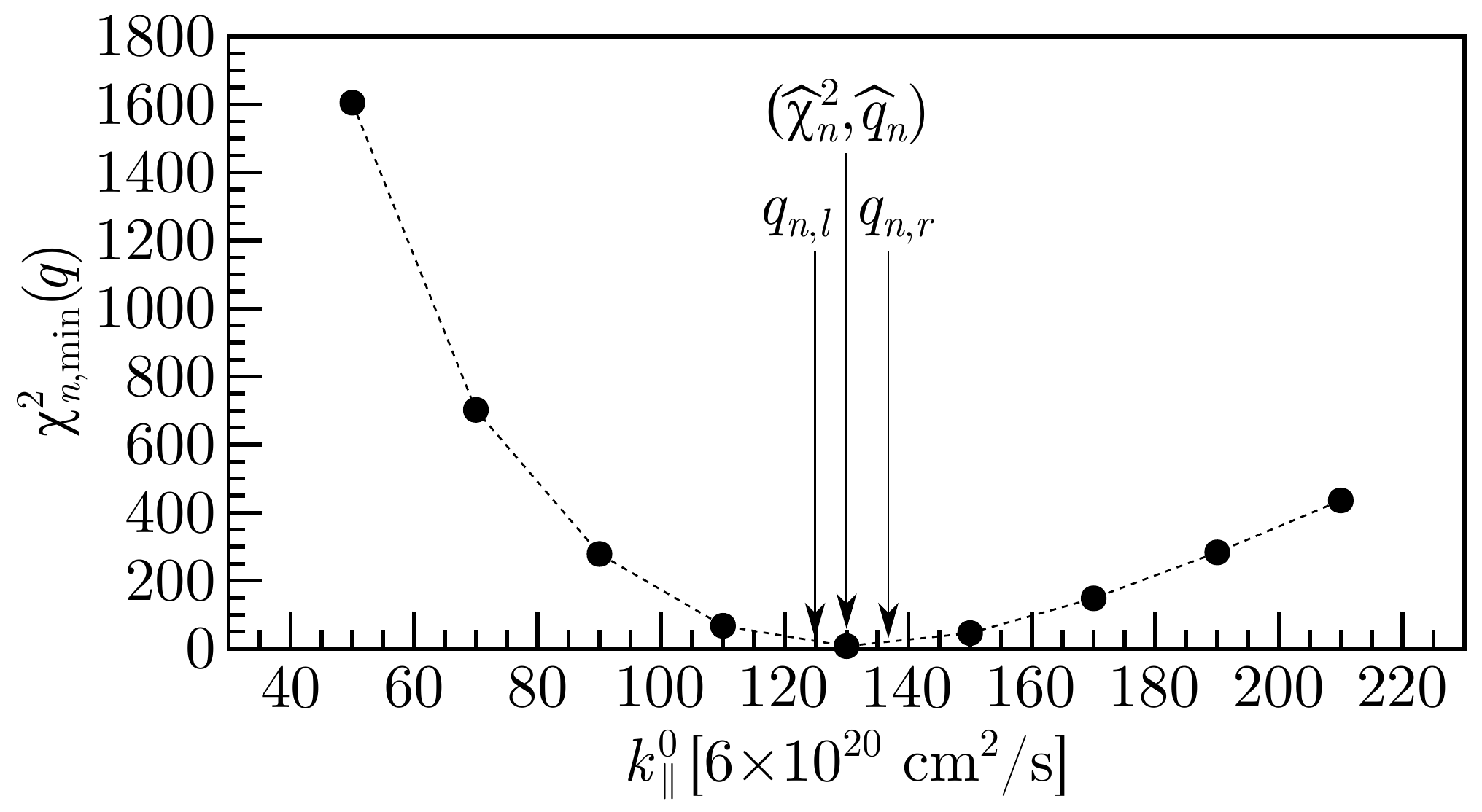}
   \caption
   {
      Uncertainty estimation for the normalization of the parallel diffusion coefficient, \cpa\ (Equation \ref{eqn:numerical-model:kpar}), in BR 2447.
      The dots are $\chi^{2}_{n,\mathrm{min}}(q)$, the position of $\widehat{\chi}^{2}_{n}$, $\widehat{q}_{n}$, $q_{n,l}$ and $q_{n,r}$ are indicated by arrows.
      The dashed line is just for guiding the eye.
   }
   \label{fig:fitting-strategy:par-chi2-scan}
\end{figure}

The uncertainty on a given parameter is estimated in the following way.
For every value $q$ of the parameter, the minimum chi-squared $\chi^{2}_{n,\mathrm{min}}(q)$ is found, regardless of the values of all the other parameters (\ie\ we marginalize over the other parameters); let's note that $\chi^{2}_{n,\mathrm{min}}(\widehat{q}_{n}) = \widehat{\chi}^{2}_{n}$, where $\widehat{q}_{n}$ is the best-fit value of the given parameter.
We then find the values $q_{n,l}$ and $q_{n,r}$, respectively to the left and right of $\widehat{q}_{n}$, for which $\chi^{2}_{n,\mathrm{min}}(q_{n,l}) = \chi^{2}_{n,\mathrm{min}}(q_{n,r}) = \widehat{\chi}^{2}_{n} + 1$.
The lower uncertainty is defined as $\widehat{q}_{n} - q_{n,l}$, while the upper uncertainty as $q_{n,r} - \widehat{q}_{n}$.
Figure \ref{fig:fitting-strategy:par-chi2-scan} shows an example of uncertainty estimation for the normalization of the parallel diffusion coefficient, \cpa, in BR 2447 (Dec. 2 -- 28, 2012), with $\widehat{\chi}^{2}_{n}$, $\widehat{q}_{n}$, $q_{n,l}$ and $q_{n,r}$ indicated by arrows.
The $\chi^{2}_{n,\mathrm{min}}(q)$ curve is well behaved, being approximately parabolic around the best-fit value.
%
%
%
\section{Numerical results} \label{sec:results}
\begin{figure*}[t]
   \centering
   \includegraphics[width=\textwidth]{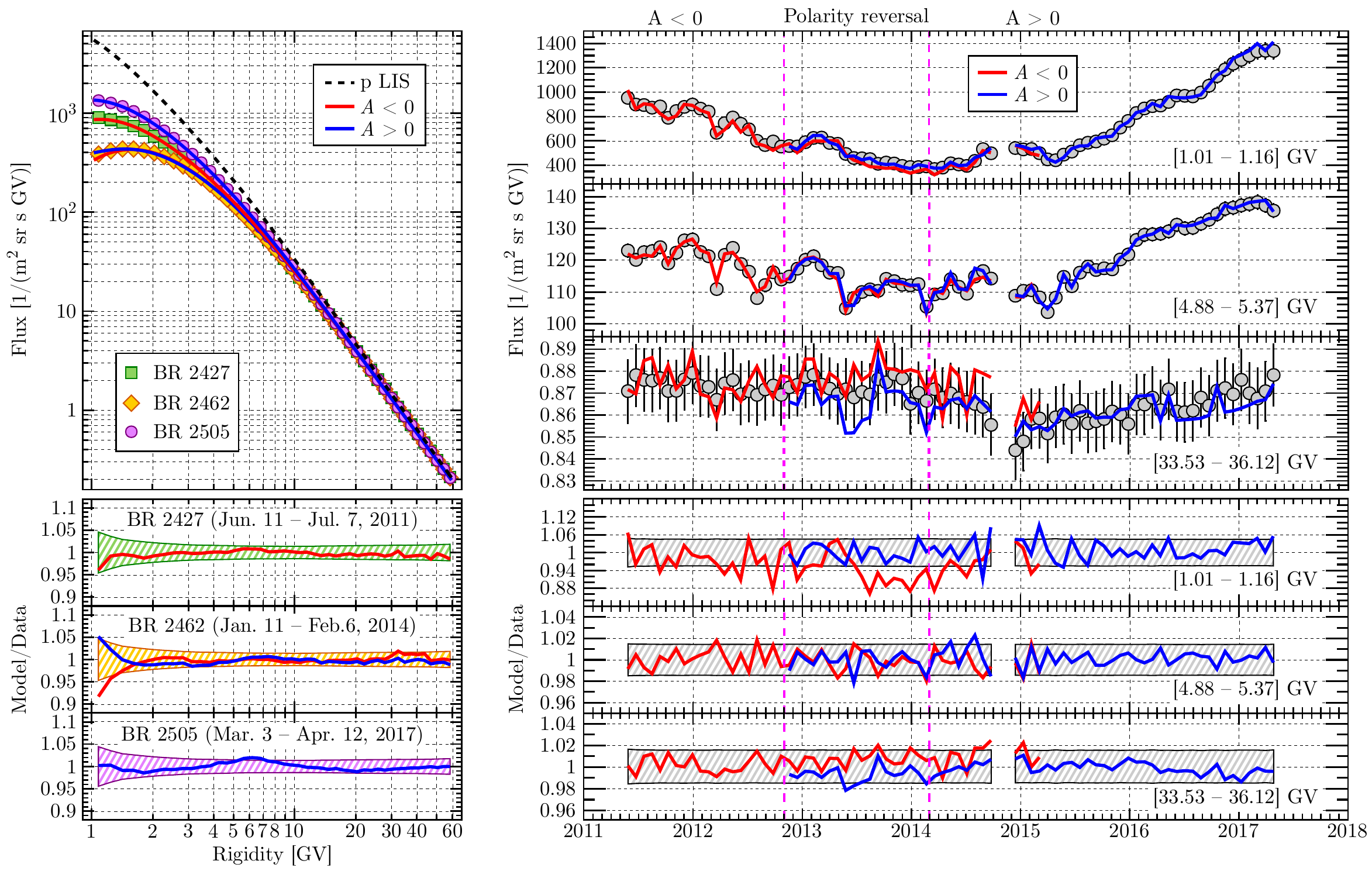}
   \caption
   {
      \textbf{Top left:} three selected AMS-02 proton fluxes (colored markers) as a function of rigidity, together with their best-fit models (red and blue lines) and the proton LIS (dashed black line).
      \textbf{Bottom left:} ratio of best-fit models to data for the three proton fluxes (red and blue lines), compared to the corresponding AMS-02 uncertainties (colored hatched bands).
      \textbf{Top right:} three selected rigidity bins of the AMS-02 proton fluxes as a function of time (gray circles), together with their best-fit models (red and blue lines).
      \textbf{Bottom right:} ratio of best-fit models to data for the three rigidity bins (red and blue lines), compared to the corresponding AMS-02 uncertainties (gray hatched bands).
      The vertical dashed magenta lines delimit the period of the solar magnetic field polarity reversal.
   }
   \label{fig:results:residuals}
\end{figure*}

Figure \ref{fig:results:residuals} shows some examples of fitted fluxes.
In the left upper panel, three AMS-02 proton fluxes at different levels of solar activity are plotted as a function of rigidity: BR 2427 (June 11 -- July 7, 2011) in green squares, corresponding to the ascending phase of solar cycle 24 and a moderate level of solar modulation; BR 2462 (January 11 -- February 6, 2014) in orange diamonds, corresponding to the solar maximum and a very depleted GCR intensity; and BR 2505 (March 3 -- April 12, 2017), in magenta circles, corresponding to the descending phase of solar cycle 24 and a low level of solar modulation.
The best-fit models are also shown: BR 2427 was modeled with negative polarity (red line), BR 2462 with both negative (red line) and positive polarity (blue line), since it was during the period of polarity reversal, and BR 2505 with positive polarity (blue line).
For reference, the proton LIS is also shown as a dashed black line.
In the left lower panel, the ratio between the best-fit models and data for the three selected fluxes (red and blue lines) is shown and compared to the corresponding uncertainty on the AMS-02 fluxes (colored hatched bands).
These plots highlight the very good agreement between the models and data at all rigidities, mostly within the experimental uncertainties.
A similar level of agreement is also obtained for all other fluxes.

In the right upper panel of Figure \ref{fig:results:residuals}, three rigidity bins of the AMS-02 proton fluxes as a function of time have been chosen (gray circles): [1.01 -- 1.16] GV, [4.88 -- 5.37] GV, and [33.53 -- 36.12] GV.
The best-fit models are shown as red (negative polarity models) and blue (positive polarity models) lines.
As previously mentioned (Section \ref{sec:fitting-strategy:heliosphere-status}), both negative and positive polarity models have been used in the period from October 2013 to February 2015.
As shown, the time dependence of the proton flux is not exactly the same at different rigidities: for example, after June 2013, the flux at 5 GV stays almost flat with month-to-month fluctuations, while the flux at 1 GV keeps decreasing until February 2014.
The flux around 35 GV, instead, is mostly constant until the maximum, then decreases of around 3.5\% over the course of 10 months after the polarity reversal, and finally starts to slowly recover (about 2\%/year) after January 2015.
All these rigidity-dependent features in the time variation of the proton fluxes are reproduced by the best-fit models.
In the right lower panel, the ratio between the best-fit models and data for the three selected rigidity bins (red and blue lines) is shown, together with the corresponding uncertainty on the AMS-02 fluxes (gray hatched bands).
The models are mostly within the experimental uncertainties at all rigidities.\\

\begin{figure*}[p]
   \centering
   \includegraphics[width=\textwidth]{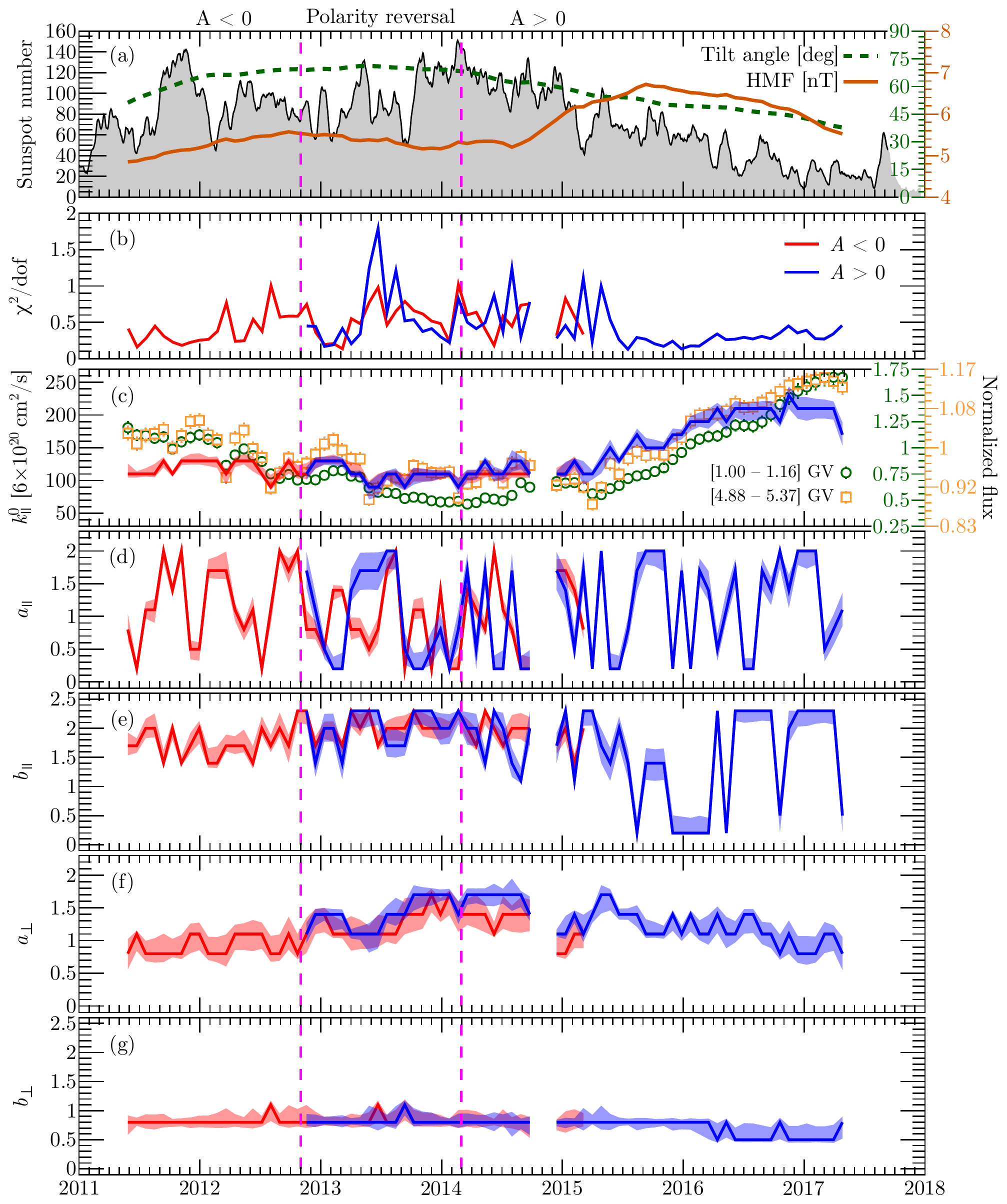}
   \caption
   {
      Time variation of the best-fit parameters (lines) and their uncertainty (bands) for models with negative (red) and positive (blue) magnetic polarity. 
      \textbf{(a)}: Sunspot number (gray area, 27-day running average), HMF (brown line) and tilt angle (dashed dark green line) used as input parameters in the models.
      \textbf{(b)}: Normalized chi-squared of the best-fit models.
      \textbf{(c)}: Normalization of the diffusion coefficient, \cpa, together with two rigidity bins of AMS-02 normalized fluxes (green circles and orange squares). 
      \textbf{(d)} and \textbf{(e)}: Low- and high-rigidity slopes of the parallel diffusion coefficient, \apar\ and \bpar.
      \textbf{(f)} and \textbf{(g)}: Low- and high-rigidity slopes of the perpendicular diffusion coefficient, \aperp\ and \bperp.
      The vertical dashed magenta lines delimit the period of the solar magnetic field polarity reversal.
   }
   \label{fig:results:model-fit-results}
\end{figure*}

The values of the best-fit parameters, together with their estimated uncertainty, are listed in Tables \ref{tab:best-fit-pars:neg} and \ref{tab:best-fit-pars:pos} in Appendix \ref{sec:best-fit-pars}.
The time variation of the best-fit parameters is analyzed in Figure \ref{fig:results:model-fit-results}.
In the top panel, the tilt angle (dashed dark green line, right axis) and the HMF (brown line, right axis) used as input in every BR are displayed for reference, together with the daily sunspot number \citepalias{bib:sidc} smoothed with a 27-day running average (gray area, left axis).

The second panel shows the normalized minimum chi-squared, $\widehat{\chi}^{2}_{n}/\mathrm{dof}$, for models with negative (red) and positive (blue) polarity.
In general, the agreement between the best-fit models and data is very good for all months, as also shown by the lower panels in Figure \ref{fig:results:residuals}.
After August 2015, the normalized chi-squared stays flatter and with fewer fluctuations with respect to the previous months: this is probably due to the fact that in this period the heliosphere is globally quieter than before and thus the steady-state approximation used to solve the Parker equation is more valid.
The sudden increases of the normalized chi-squared for the positive polarity models in the middle of 2013, during the period of the polarity reversal, might be considered as statistical fluctuations, but also as an indication that modeling a mixed-polarity heliosphere is necessary to correctly describe GCR fluxes during the solar maximum.

The third panel shows the best-fit values (lines) for the normalization of the parallel diffusion coefficient (Equation \ref{eqn:numerical-model:kpar}) with their estimated uncertainty (bands), together with the monthly AMS-02 fluxes in the rigidity bins [1.00 -- 1.16] GV (green circles) and [4.88 -- 5.37] GV (orange squares), respectively normalized to their averaged values.
The variations of \cpa\ follow closely the time dependence of the observed fluxes (especially around 5 GV), as expected, since \cpa\ is the main parameter that controls the level of modulation.
For example, the drops of \cpa\ (\ie\ short-term increases in the modulation strength) correspond with the drops of the proton fluxes, \eg\ in October 2011 or March 2012.
A caveat of this analysis is that these drops are due to CMEs hitting the Earth, \ie\ local disturbances, which are not included in the model.
Nevertheless, the model is able to reproduce the flux, by changing globally the diffusion coefficient in order to match the local conditions: we expect that, in these cases, the solution at positions far from Earth will not be accurate, since the diffusion in these positions is not affected by the CME.
It is worth noting that, in the period of the polarity reversal, the best-fit \cpa\ obtained from models with negative polarity agrees with the one from models with positive polarity, \ie\ the normalization of the diffusion coefficient seems to be mostly insensitive to the sign of the HMF polarity.
We computed the Pearson correlation between \cpa\ and the proton flux intensity at different rigidities, taking into account the uncertainties on the measured fluxes and on the best-fit values with a toy Monte Carlo.
The maximum correlation $r = 0.82$, with a 95\% confidence interval of (0.78, 0.85), is found around 5 GV, while at 1 GV $r = 0.73$, with a 95\% confidence interval of (0.68, 0.77).
The correlation becomes consistent with 0 at the 95\% confidence level around 22 GV.

Panels (d) and (e) show the time variation of the low- and high-rigidity slope of the parallel diffusion coefficient, \apar\ and \bpar.
The best-fit values considerably vary from month to month, making it difficult to discern any clear time-dependent pattern.
Indeed, sometimes the $\chi^{2}_{min}(q)$ curve has two local minima or it does not have a parabolic behavior.
This means that these two parameters are not well constrained by fitting the AMS-02 proton fluxes, implying that the modulated flux is not so sensitive to the values of \apar\ and \bpar\ for rigidities above 1 GV.
A possible explanation is that the parallel diffusion dominates very close to the Sun, when most of the modulation has already happened.
The diffusion coefficient in the radial direction is $\Krr = k_{\parallel}\cos^{2}\psi + k_{\perp,r}\sin^{2}\psi$; imposing equality between the two terms yields $\tan^{2}\psi \simeq 1/k_{\perp,r}^{0} = 50$, corresponding to a spiral angle $\psi \approx 80 \degree$, which can be found already around 5 AU, a mere 0.01\% of the whole heliosphere volume.
The time variation of GCR protons measured by PAMELA down to 400 MV \citep{bib:adriani13:pamela-p-monthly-flux,bib:adriani18:pamela-p-monthly-flux} would provide a better constraints on the slopes of the parallel diffusion coefficient; this study will be the focus of a future work.

The parameters describing the perpendicular diffusion coefficient, \aperp\ and \bperp, are shown in the panels (f) and (g).
Remarkably, \bperp\ is almost constant with time, both for positive and negative polarity, whose best-fit values agree in almost all the overlapping months.
\aperp\ is mostly flat before the maximum of solar activity, when $A<0$.
During the period of the polarity reversal \aperp\ rises, almost doubling its value (with respect to 2011 and 2012) as the solar activity peaks, showing an anti-correlation with the proton flux at 1 GV (see the third panel).
This suggests that, on top of the overall modulation scale determined by \cpa, low rigidities experience an even smaller perpendicular mean free path.
This is also supported by computing the Pearson correlation between \aperp\ and the proton flux intensity at different rigidities: the maximum anti-correlation $r = -0.5$, with a 95\% confidence interval of $(-0.62,\, -0.39)$, is found at 1 GV, while it decreases with increasing rigidity, becoming consistent with 0 above 20 GV.
As for \cpa, during the period of the polarity reversal the best-fit \aperp\ and \bperp\ obtained from models with negative polarity agree, within the fit uncertainties, with the ones from models with positive polarity.

We verified that these results do not depend on the duration of the period used to compute the backward average of the HMF and the tilt angle (see Section \ref{sec:fitting-strategy:heliosphere-status}).
We varied the number of months ($n$) included in the average between 0 and 24 months, in steps of 2 months.
For $n \geq 4$ months, the values of the best-fit parameters are consistent, within uncertainties, with the ones presented in Figure \ref{fig:results:model-fit-results}, while the residuals between the best-fit models and the data are similar to the ones shown in Figure \ref{fig:results:residuals}.
For $n = 0$ and $n = 2$ months, the normalized chi-squared and the residuals are worse in a few BRs between the end of 2014 and the beginning of 2015, when the HMF has a higher variability than in the rest of the analyzed period.
This suggests that the steady-state approximation is a valid approach to describe the time variation of GCRs above 1 GV on a monthly basis, provided the heliosphere status is adjusted by smoothing the input HMF and tilt angle with a backward average of at least 4 months.
%
%
%
\section{p/He ratio comparison} \label{sec:phe-comparison}
It is generally assumed in modulation studies that the rigidity dependence of the three mean free paths is the same for all nuclei.
The assumption has not been rigorously tested because the observational data were never accurate enough over the relevant rigidity range for all cosmic ray nuclei over a complete solar cycle.
Under this assumption, the best-fit parameters derived in Section \ref{sec:results} from AMS-02 protons should also be valid for other nuclei, in particular \He3\ and \He4.
In order to compute the modeled p/He ratio, we ran the best-fit models for \He3\ and \He4\ (with the corresponding charge, mass and LIS), and then we summed the resulting modulated fluxes.

\begin{figure*}[t]
   \centering
   \includegraphics[width=\textwidth]{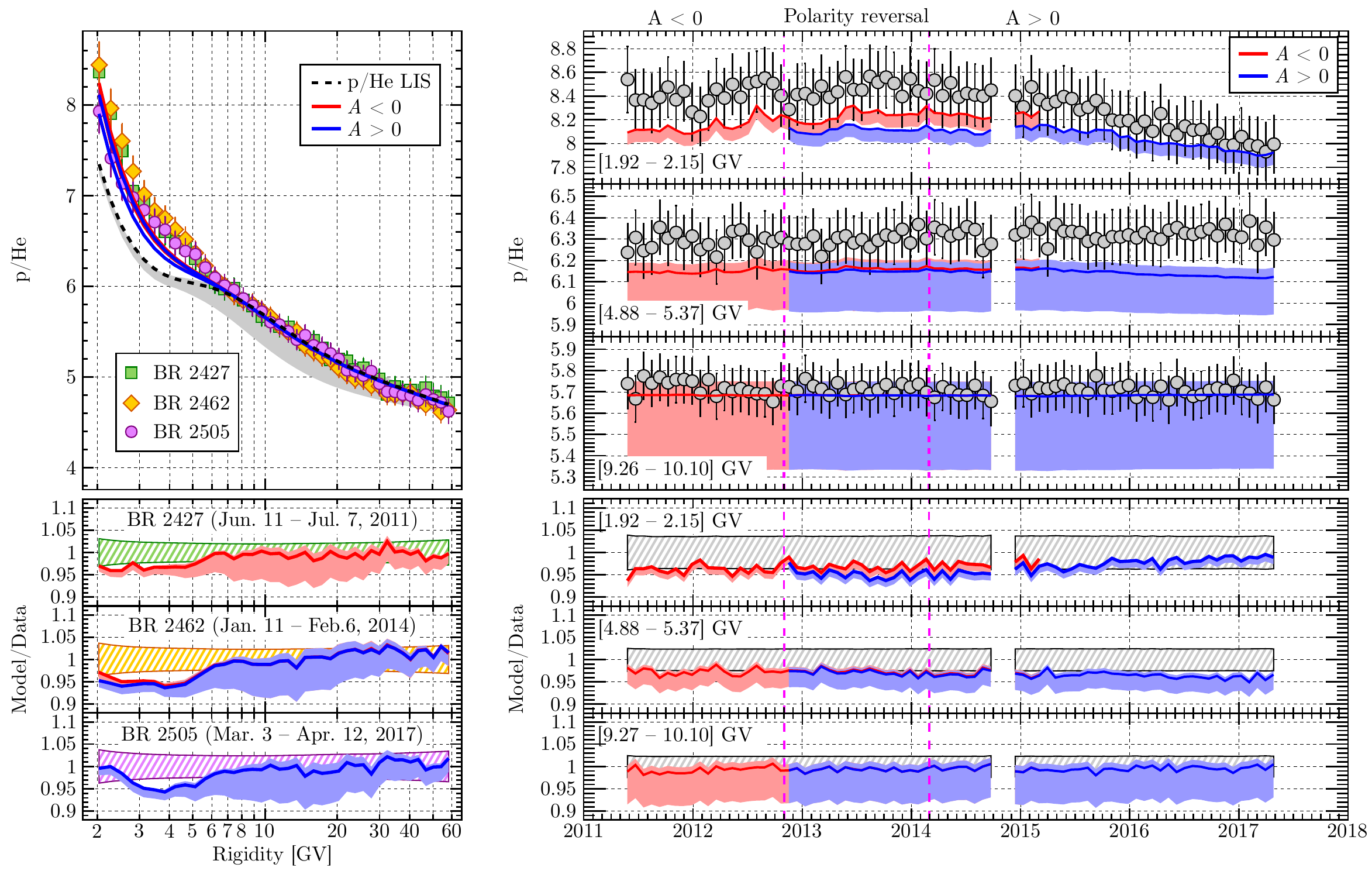}
   \caption
   {
      \textbf{Top left:} three selected AMS-02 p/He ratios (colored markers) as a function of rigidity, together with their best-fit models (red and blue lines) and the p/He LIS (dashed black line).
      \textbf{Bottom left:} ratio of best-fit models to data for the three BRs (red and blue lines), compared to the corresponding AMS-02 uncertainties (green, orange and magenta hatched bands).
      Light red, blue and gray bands represent the p/He uncertainty due to different \He3\ and \He4\ LIS parametrizations.
      \textbf{Top right:} three selected rigidity bins of the AMS-02 p/He ratio as a function of time (gray circles), together with their best-fit models (red and blue lines and bands).
      \textbf{Bottom right:} ratio of best-fit models to data for the three rigidity bins (red and blue lines and bands), compared to the corresponding AMS-02 uncertainties (gray hatched bands).
      The vertical dashed magenta lines delimit the period of the solar magnetic field polarity reversal.
   }
   \label{fig:phe-comparison:residuals}
\end{figure*}
Figure \ref{fig:phe-comparison:residuals} shows the comparison between the p/He ratio observed by AMS-02 and the one predicted by the model.
In the left upper panel, three p/He AMS-02 ratios for the same BRs from Figure \ref{fig:results:residuals} are plotted as a function of rigidity (colored markers), together with the best-fit models (blue and red lines, respectively for negative and positive polarity models).
For reference, the p/He LIS is also shown as a dashed black line.
In the left lower panel, the ratio between the best-fit models and data for the three selected BRs (red and blue lines) is shown and compared to the corresponding uncertainty on the AMS-02 p/He ratios (green, orange and magenta hatched bands).
The light red, blue and gray bands represent the uncertainty on p/He due to the uncertainty on the \He3\ and \He4\ LIS.
This uncertainty has been estimated by varying the datasets used to derive the \He3\ and \He4\ LIS and by assuming (or not) the same modulation potential for \He3\ and \He4\ (see Section \ref{sec:numerical-model:LIS}).
A total of 16 different LIS parametrizations have been computed, and for each of them the best-fit models have been run.
The minimum and maximum value among the different parametrizations at each rigidity has been considered as the uncertainty on the modulated p/He.
While the difference in LIS parametrizations above 2 GV is between 10\% and 60\% for \He3, and between 5\% and 10\% for \He4, the uncertainty on the modulated p/He is relatively small, less than 4\%.
For comparison, the uncertainty on the proton LIS coming from the parametrization fit is less than 2\%, so its contribution to the modulated p/He uncertainty is considered negligible.
In the following, the LIS parametrization described in Section \ref{sec:numerical-model:LIS} will be called \emph{reference LIS}.

In the right upper panel of Figure \ref{fig:phe-comparison:residuals}, three rigidity bins of the AMS-02 p/He ratio as a function of time have been chosen (gray circles): [1.92 -- 2.15] GV, [4.88 -- 5.37] GV, and [9.26 -- 10.10] GV.
The best-fit models, together with their uncertainty, are shown as red (negative polarity models) and blue (positive polarity models) lines and bands.
In the right lower panel, the ratio between the best-fit models and data for the three selected rigidity bins (red and blue lines and bands) is shown, together with the corresponding uncertainty on the AMS-02 fluxes (gray hatched bands).
We can see that the modeled p/He using the reference LIS on average underestimates the data by 5\% below 6 GV.
This difference remains basically constant in time, amounting to a rigidity-dependent normalization shift in the modulated p/He below 6 GV, which persists even considering the modeled p/He uncertainties.
Indeed, the different LIS parametrizations result in similar p/He time variations, differing only for a shift constant in time.
This might be due to two reasons: (a) the \He3\ and \He4\ LIS parametrizations are not correct below 5 GV; (b) the assumption of same mean free path for p and He at all relevant rigidities is inadequate.
We believe that (a) is the most probable explanation: indeed, the use of the force-field approximation to derive the LIS might introduce a bias in the resulting parametrization, which could affect the results of the numerical model analysis.
%
%
%
\section{Time dependence of p/He} \label{sec:phe-time-dep}
AMS-02 data shows that above 3 GV the p/He ratio is time independent.
Below 3 GV, it is flat within month to month variations until March 2015, and then it starts to decrease, seemingly correlated with the decrease in solar activity.
As stated in \cite{bib:aguilar18:ams-monthly-phe}, the origin of the p/He time dependence may be due to: (a) the difference in LIS shape between p and He; (b) the dependence of the diffusion tensor on the particle mass-to-charge ratio, $A/Z$.
For the sake of simplicity, let's examine the steady-state one dimensional version of the Parker equation:
\begin{equation} \label{eqn:phe-time-dep:parker}
V \frac{\partial f}{\partial r} - \frac{1}{r^{2}}\frac{\partial}{\partial r} \left( k r^{2} \frac{\partial f}{\partial r} \right) - \frac{R}{3r^{2}} \frac{\partial}{\partial r} \left( r^{2} V \right) \frac{\partial f}{\partial R} = 0,
\end{equation}
where $V$ is the solar wind speed and $k = k(r,R,A/Z) = \frac{1}{3} v(R,A/Z) \lambda(r,R)$ is the radial diffusion coefficient.
$v(R,A/Z) = \beta c = c/\sqrt{1+(A/Z)^{2}(mc/eR)^{2}}$ is the particle velocity, with $m$ the proton rest mass, while $\lambda(r,R)$ is the mean free path, which is assumed to depend only on the radial distance and rigidity, \ie\ to be the same for all nuclei.

Let's assume that two nuclei, $p_{1}$ and $p_{2}$, have the same $A/Z$, but a different LIS shape: then the diffusion coefficient will be the same for both, but the boundary conditions will not be equal.
In particular, the last term is sensitive to the spectral index, $\Gamma = \partial \log(f)/\partial \log(R) = R/f\; \partial f/\partial R$, so we can expect that the difference in $\Gamma$ at the heliopause will persist during the propagation through the heliosphere.
Since $V$ and $k$ change with varying modulation conditions, $\Gamma(p_{1}/p_{2}) = \Gamma(p_{1}) - \Gamma(p_{2})$ will be changing as well, \ie\ the ratio $p_{1}/p_{2}$ at a given rigidity will not be constant in time.
Now, let's assume that two nuclei with different $A/Z$ have the same LIS shape: then all the terms in the Parker equation are the same for the two species, except for the divergence of the diffusive flux, because of the $A/Z$ dependence of $k$.
A time variation of $k$ will translate into a time variation of $p_{1}/p_{2}$ at a given rigidity.
It is important to note that this dependence comes from the fact that we assumed $\lambda$ to depend only on $R$: if $\lambda$ was also a function of $A$ and $Z$, the $A/Z$ dependence of $v\lambda$ might cancel out.
Effectively, we can say that if two nuclei with different $A/Z$ have the same mean free path, then the time variation of $p_{1}/p_{2}$ is a natural consequence, even when the LIS shape is the same.

The same reasoning can be applied to the full 3D case: the symmetric components of the field-aligned diffusion tensor $\mathbf{K}$ can all be written as $k_{i} = \frac{1}{3}v\lambda_{i}$, where $i$ stands for the parallel, perpendicular radial and perpendicular polar directions, so that also $\mathbf{K}$ depends explicitly on $A/Z$.

In the case of p/He, all the species involved have different $A/Z$ and different LIS, so the time dependence is due to a combination of (a) and (b).
To assess which of the two causes is dominant, we test separately the effect of (a) and (b).
Since the uncertainties on the \He3\ and \He4\ LIS parametrizations does not affect the modeled p/He time dependence at a given rigidity, but only its normalization, in the rest of this section we will use the reference LIS and we will compare the normalized modeled p/He to the normalized observed p/He, so to remove any normalization shift.
We verified that the uncertainty on the normalized p/He due to the uncertainty in the \He3 and \He4 LIS is less than 0.5\% at all rigidities and for each BR.

\subsection{Difference in the LIS shape} \label{sec:phe-time-dep:lis}
To understand the effect of the difference in the LIS shape, we ran the best-fit models for p, \He3, and \He4, forcing the same $A/Z$ for all three species, but using the appropriate LIS for each particle.
In the following, p corresponds to the proton LIS and $A/Z=1$, \He3\ corresponds to the \He3\ LIS and $A/Z=1$, while \He4\ corresponds to the \He4\ LIS and $A/Z=1$.
The same results are obtained if we use $A/Z=3/2$ or $A/Z=2$ for all particles.

\begin{figure*}[t]
   \centering
   \includegraphics[width=\textwidth]{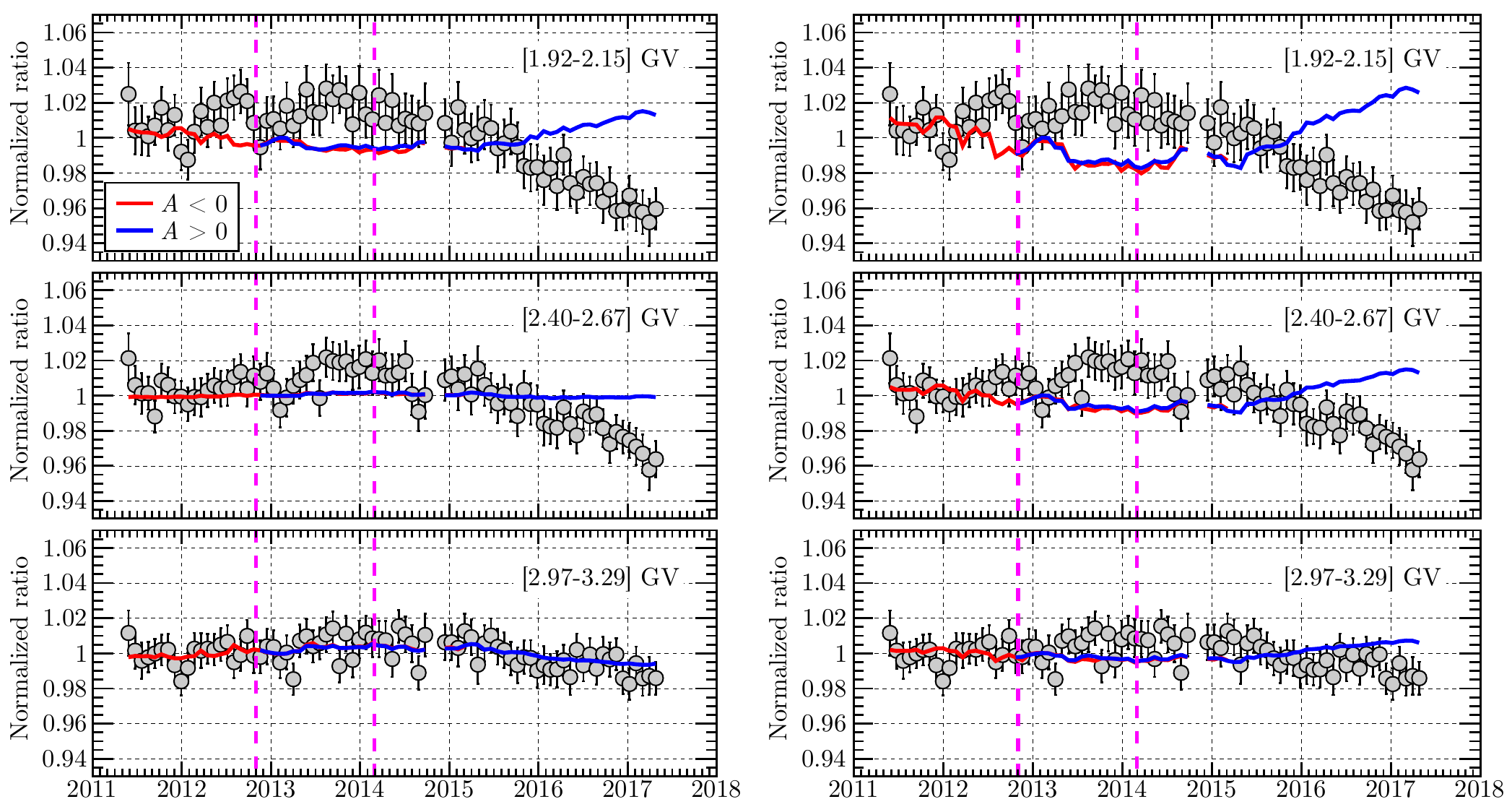}
   \caption
   {
      Effect of the difference in LIS shape on the time variation of p/He.
      Normalized modeled p/\He3 (red and blue lines, left) and p/\He4 (red and blue lines, right) compared to the observed p/He (gray circles) as a function of time for three selected rigidity bins.
      The vertical dashed magenta lines delimit the period of the solar magnetic field polarity reversal.
   }
   \label{fig:phe-time-dep:lis}
\end{figure*}

Figure \ref{fig:phe-time-dep:lis} shows the comparison of the normalized modeled p/\He3\ (red and blue lines, left panels) and p/\He4\ (red and blue lines, right panels) with the normalized observed p/He (gray circles) for three selected rigidity bins: [1.92 -- 2.15] GV, [2.40 -- 2.67] GV, and [2.97 -- 3.29] GV.
Note that the plotted experimental uncertainties are the sum in quadrature of the statistical and time-dependent uncertainties only: the systematic uncertainties constant in time are not considered here, since they would affect only the average value of p/He in a given rigidity bin, but not its time variation.\\
The time trend of the observed p/He is not reproduced: p/\He4\ increases with time at all rigidities after March 2015, while p/\He3\ increases at 2 GV, stays flat at 2.5 GV, and slightly decreases at 3 GV.

We can better understand the different behavior of p/\He3\ and p/\He4\ by looking at the spectral index, $\Gamma$, of the LIS ratio.
Because of the adiabatic energy losses, the observed particles at 2 GV had a greater rigidity before entering the heliosphere, so in order to compare $\Gamma$ in interstellar space we should correct for this effect.
In the force-field approximation framework, the energy losses are related to the modulation potential $\phi$, whose values usually vary between few hundreds of MV and 1 GV, depending on the level of solar activity.
Using $\phi = 400$ MV as the average modulation potential in the descending phase of the solar cycle, we can relate the rigidity observed at Earth $R_{E} = \sqrt{T_{E}(T_{E}+2Amc^{2})}/Ze$ with the rigidity at the heliopause $R_{HP} = \sqrt{T_{HP}(T_{HP}+2Amc^{2})}/Ze$, where $T_{HP} = T_{E} + Ze\phi$, while   $T_{E}$ and $T_{HP}$ are, respectively, the kinetic energy at Earth and at the heliopause.
With this choice, we find that at 2 GV, $\Gamma(\p/\He3) = -0.25$, while $\Gamma(\p/\He4) = -0.39$; at 2.5 GV, $\Gamma(\p/\He3) = -0.03$, while $\Gamma(\p/\He4) = -0.26$; at 3 GV, $\Gamma(\p/\He3) = 0.12$, while $\Gamma(\p/\He4) = -0.16$.
Note that when the values of $\Gamma(\p/\He3)$ and $\Gamma(\p/\He4)$ are very similar in absolute value, so is the amplitude of the time variation in the normalized modeled ratios.
A different choice of $\phi$ leads to different values for $\Gamma$, but qualitatively the comparison remains the same: $\Gamma(\p/\He4)$ is always negative and decreases in absolute value with increasing rigidity, while $\Gamma(\p/\He3)$ is negative at 2 GV, very close to 0 at 2.5 GV, and positive at 3 GV.
This suggests that the time behavior of the ratio of two species with the same $A/Z$ is related to the spectral index of the LIS ratio of the two species: if $\Gamma < 0$, then the ratio will be anti-correlated with the solar activity, while if $\Gamma > 0$, the ratio will be correlated with the solar activity.
The amplitude of the time variation is instead proportional to the absolute value of $\Gamma$.
We verified that this result holds when considering different parametrizations for the \He3 and \He4 LIS.
The uncertainties on the \He4 LIS are small enough that $\Gamma(\p/\He4)$ is always negative, leading to an increase of p/\He4.
Instead, the uncertainties on the \He3 LIS are such that $\Gamma(\p/\He3)$ can be positive or negative depending on the parametrization, and so p/\He3 decreases or increases with time according to the sign of $\Gamma(\p/\He3)$.
Since \He4 accounts for around 80\% of the He, the p/He behaviour is dominated by p/\He4, and thus, even taking into account the uncertainty on the \He3 and \He4 LIS, the observed p/He can not be reproduced if we assume the same $A/Z$, but different LIS.

The relation between the time variation and the spectral index could be tested with a long-term measurement of the ratio of two species with exactly the same $A/Z$, for example deuterons and \He4.
AMS-02 might be able to perform such a measurement, because of its large acceptance and precision.

\subsection{Charge-to-mass ratio dependence of the diffusion tensor} \label{sec:phe-time-dep:velocity}
\begin{figure*}[t]
   \centering
   \includegraphics[width=\textwidth]{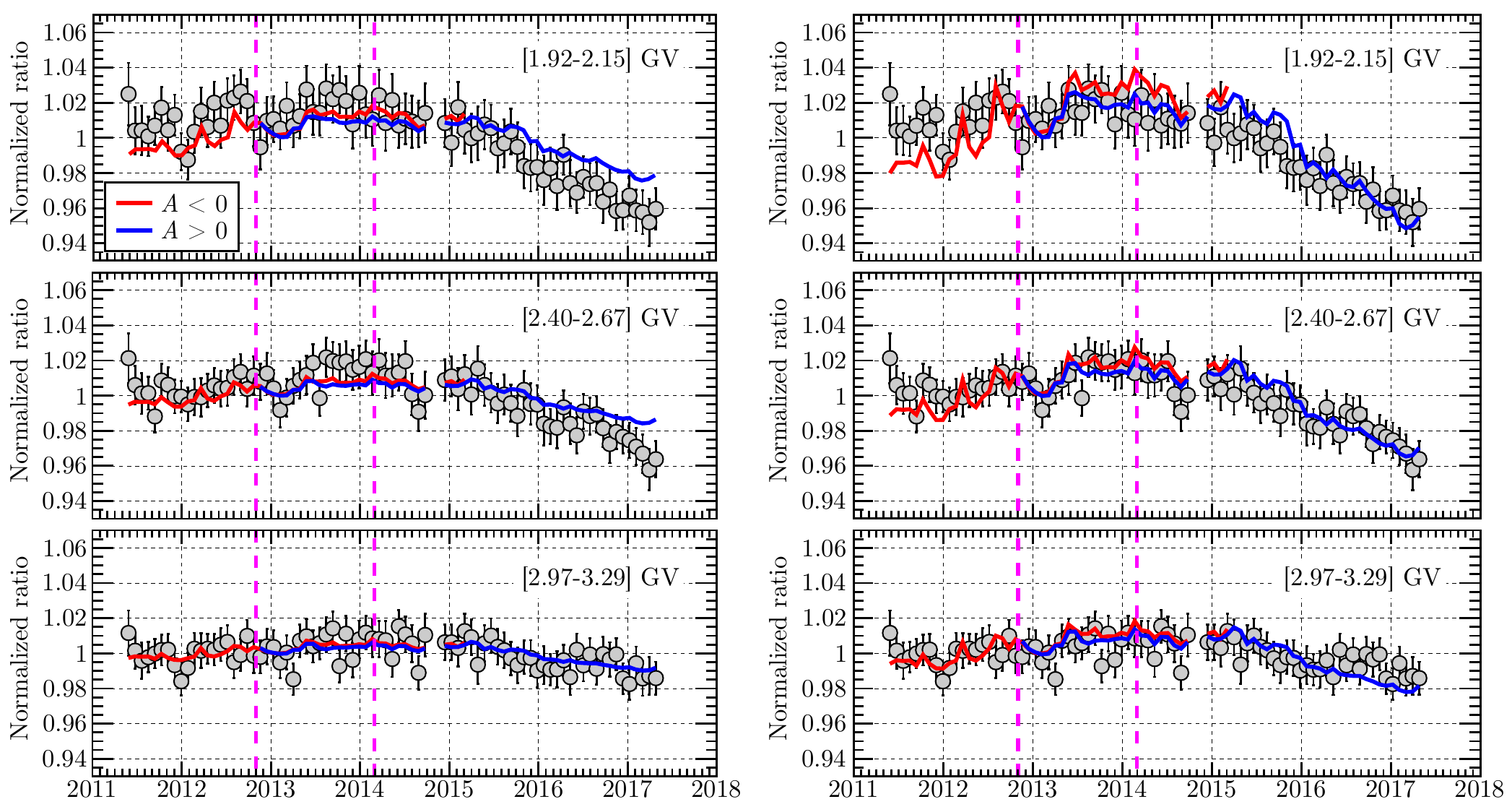}
   \caption
   {
      Effect of the $A/Z$ dependence of the diffusion tensor on the time variation of p/He.
      Normalized modeled p/\He3 (red and blue lines, left) and p/\He4 (red and blue lines, right) compared to the observed p/He (gray circles) as a function of time for three selected rigidity bins.
      The vertical dashed magenta lines delimit the period of the solar magnetic field polarity reversal.
   }
   \label{fig:phe-time-dep:velocity}
\end{figure*}
To understand the effect of the $A/Z$ dependence of the diffusion tensor, we ran the best-fit models for p, \He3, and \He4, forcing the same LIS for all three species, but using the appropriate $A/Z$ for each particle.
In the following, p corresponds to the proton LIS and $A/Z=1$, \He3\ corresponds to the proton LIS and $A/Z=3/2$, while \He4\ corresponds to the proton LIS and $A/Z=2$.

Figure \ref{fig:phe-time-dep:velocity} shows the comparison of the normalized modeled p/\He3\ (red and blue lines, left panels) and p/\He4\ (red and blue lines, right panels) with the normalized observed p/He (gray circles) for the same rigidity bins as in Figure \ref{fig:phe-time-dep:lis}.\\
The time trend of the observed p/He is reproduced, but for \He3\ models the amplitude of the decrease below 3 GV is smaller than for data and \He4\ models: the difference in $A/Z$ between \He3\ and \He4\ is clearly playing an important role.

It is interesting to consider the ratio between particle velocities: at 2 GV, $v(\p)/v(\He3) = 1.1$, while $v(\p)/v(\He4) = 1.23$; at 2.5 GV, $v(\p)/v(\He3) = 1.07$, while $v(\p)/v(\He4) = 1.16$; at 3 GV, $v(\p)/v(\He3) = 1.05$, while $v(\p)/v(\He4) = 1.12$.
The values of $v(\p)/v(\He3)$ at 2 GV and $v(\p)/v(\He4)$ at 3 GV are very similar, and so is the amplitude of the decrease of p/\He3\ at 2 GV and p/\He4 at 3 GV.
This suggests that the magnitude of the time variation of the ratio of two species with the same LIS is proportional to the velocity ratio of the two species.
As seen in Section \ref{sec:results}, \cpa\ is the main parameter that determines the level of modulation of the flux, so it's reasonable to expect that the amplitude of the variation of p/He depends on the component-wise ratio $\mathbf{K}(\p)/\mathbf{K}(\He{}) = v(\p)/v(\He{})\mathds{1}$.

These results do not depend on the choice of the LIS.
Using any parametrization of the \He3\ or \He4\ LIS we obtain the same normalized p/\He3\ and p/\He4\ as the results presented in Figure \ref{fig:phe-time-dep:velocity}.\\

The two tests show that the observed p/He time variation is most probably due to the $A/Z$ dependence of the diffusion tensor, since the difference in LIS shape should produce the opposite time behavior between 2 and 3 GV.
However, PAMELA is able to measure p/He at lower rigidities than AMS-02 (down to 0.4 GV) and took measurements at different solar activity conditions (from the minimum of solar cycle 23/24 to the maximum of solar cycle 24).
At 1 GV, the spectral index of p/\He3\ and p/\He4\ LIS ratio is negative and greater (in absolute value) than at 2 GV, so it might be possible that the difference in LIS shape plays a bigger role at 1 GV than at 2 GV.

As noted before, the symmetric components of $\mathbf{K}$ depend on $A/Z$, but the diffusion tensor also contains the drift coefficient, $k_{A}$, in its anti-symmetric part.
From Equation \ref{eqn:numerical-model:drift}, we see that $k_{A}$ is a function of $A/Z$, so we might expect a different behavior of the time dependence of p/He according to the HMF polarity.
We verified that this is not the case, by repeating the tests with best-fit models with both polarities for all BRs.
The difference between $A<0$ and $A>0$ models is of the order of 0.5\% for $R \geqslant 2$ GV.
The drift effects for protons becomes larger below 1 GV \citep{bib:potgieter17:diff-mod-pro-ele}, so PAMELA data might be able to reveal a difference in the time variation of p/He before and after the polarity reversal, due to the $A/Z$ dependence of the drift coefficient.
%
%
%
\section{Conclusions} \label{sec:conclusions}
Understanding diffusion processes in the heliosphere is crucial to improve predictions of the time variation of GCR fluxes at Earth and other locations of interest.
The recently published monthly proton and helium fluxes measured by AMS-02 allow the detailed study of the effects of solar modulation, during the ascending phase, solar maximum, and descending phase of solar cycle 24.
AMS-02 observed that the proton flux at 1 GV and at 5 GV behaves differently with time during the period of the solar maximum: the flux intensity at 1 GV keeps decreasing until the peak of solar activity, while at 5 GV remains flatter.
Instead, the AMS-02 p/He ratio below 3 GV is constant in time until March 2015, then it starts to decrease, at the same time that proton and helium fluxes start to recover, while the solar cycle progresses toward the next minimum.

In this work, a sophisticated state-of-the-art 3D numerical model has been tuned to reproduce the monthly proton fluxes measured by AMS-02. The fitted normalization of the parallel diffusion coefficient is well correlated with the proton flux intensity at 5 GV and anti-correlated with the sunspot number.
The different time behavior of the proton flux at 1 GV and 5 GV is determined by the slope of the perpendicular diffusion coefficient.
During the period of the maximum solar activity, the perpendicular mean free path decreases more at low rigidities than at high rigidities.

Assuming the same mean free path for p, \He3 and \He4, the best-fit models are able to reproduce the observed time trend of the p/He ratio, albeit with a small rigidity-dependent normalization shift, most probably due to a bias in the \He3 and \He4 LIS parametrization.

To understand the origin of the time dependence of p/He, two separate tests were performed.
First, the model was run assuming a different LIS for p, \He3 and \He4, but the same mass-to-charge ratio, $A/Z$, to explore how the difference in LIS shape can affect the time variation of p/He.
Then, the model was run assuming the same LIS for p, \He3 and \He4, but a different $A/Z$, to check the effect of the $A/Z$ dependence of the diffusion tensor.
The second test was able to reproduce the observed p/He time variation, while the first test was not.
Thus, the $A/Z$ dependence of the diffusion tensor seems to be the dominant cause of the time variation of p/He, at least in the rigidity range between 2 GV and 3 GV.
Data from PAMELA on p/He at lower rigidities and from AMS-02 on d/\He4 would shed light on the importance of the difference in the LIS shape.

\acknowledgments
\emph{Note added} -- While this work was in review, we became aware of a related study from \cite{bib:tomassetti18:ams}.
Their work is based on the same data sets of ours, but it follows different approaches for the galactic and heliospheric transport modeling.
Their results are consistent with those presented in this manuscript.\\

We would like to thank E. E. Vos, R. D. Strauss, N. Tomassetti, and S. Della Torre for fruitful discussions about the physics of GCR transport in the heliosphere.
We acknowledge the financial support of: National Science Foundation Early Career under grant (NSF AGS-1455202); Wyle Laboratories, Inc. under grant (NAS 9-02078); NASA under grant (17-SDMSS17-0012).
MSP acknowledges the financial support of the South African National Research Foundation (NRF) under the Competitive Funding for Rated Researchers, grant 68198.

\appendix

\section{Best-fit parameters} \label{sec:best-fit-pars}

\startlongtable
\begin{deluxetable*}{c|cc|CCCCCCC}
   \tablecaption{
      Best-fit parameters used as input for numerical models with negative polarity. \label{tab:best-fit-pars:neg}
   }
   \tablehead{
      \colhead{BR\tablenotemark{a}} & \colhead{\tilt\tablenotemark{b}} & \colhead{\cmf\tablenotemark{c}} & 
      \colhead{\cpa\tablenotemark{d}} & \colhead{$\lambda_{\perp}$(1 GV)\tablenotemark{e}} & \colhead{$\lambda_{\perp}$(5 GV)\tablenotemark{e}} & \colhead{\apar\tablenotemark{f}} & \colhead{\bpar\tablenotemark{g}} & 
      \colhead{\aperp\tablenotemark{h}} & \colhead{\bperp\tablenotemark{i}}
   }
   \startdata
2426  &  51.20  &  4.85  &  110^{+20}_{-5} &  0.009^{+0.003}_{-0.003}    &  0.034^{+0.005}_{-0.004}    &  0.8^{+0.2}_{-0.2}    &  1.7^{+0.2}_{-0.2}   &  0.8^{+0.2}_{-0.2}  &  0.8^{+0.1}_{-0.2}   \\
2427  &  53.55  &  4.87  &  110^{+4}_{-5}  &  0.006^{+0.0005}_{-0.002}   &  0.032^{+0.002}_{-0.003}    &  \leqslant 0.3        &  1.7^{+0.1}_{-0.1}   &  1.1^{+0.05}_{-0.3} &  0.8^{+0.08}_{-0.06} \\
2428  &  55.33  &  4.93  &  110^{+9}_{-4}  &  0.009^{+0.003}_{-0.002}    &  0.034^{+0.004}_{-0.002}    &  1.1^{+0.1}_{-0.1}    &  2^{+0.2}_{-0.2}     &  0.8^{+0.2}_{-0.1}  &  0.8^{+0.1}_{-0.1}   \\
2429  &  57.05  &  4.96  &  110^{+20}_{-5} &  0.009^{+0.003}_{-0.002}    &  0.033^{+0.005}_{-0.003}    &  1.1^{+0.2}_{-0.2}    &  2^{+0.3}_{-0.2}     &  0.8^{+0.2}_{-0.1}  &  0.8^{+0.1}_{-0.1}   \\
2430  &  58.67  &  5.05  &  130^{+6}_{-10} &  0.011^{+0.005}_{-0.002}    &  0.039^{+0.004}_{-0.004}    &  \geqslant 1.8        &  1.4^{+0.2}_{-0.07}  &  0.8^{+0.3}_{-0.1}  &  0.8^{+0.1}_{-0.09}  \\
2431  &  60.34  &  5.10  &  110^{+5}_{-4}  &  0.009^{+0.003}_{-0.0008}   &  0.032^{+0.003}_{-0.002}    &  1.4^{+0.1}_{-0.1}    &  2^{+0.1}_{-0.2}     &  0.8^{+0.2}_{-0.05} &  0.8^{+0.06}_{-0.08} \\
2432  &  62.13  &  5.13  &  130^{+4}_{-7}  &  0.011^{+0.004}_{-0.001}    &  0.038^{+0.003}_{-0.002}    &  \geqslant 1.9        &  1.4^{+0.1}_{-0.04}  &  0.8^{+0.3}_{-0.08} &  0.8^{+0.07}_{-0.07} \\
2433  &  63.64  &  5.14  &  130^{+10}_{-5} &  0.007^{+0.002}_{-0.003}    &  0.035^{+0.003}_{-0.003}    &  0.5^{+0.1}_{-0.2}    &  1.7^{+0.2}_{-0.1}   &  1.1^{+0.2}_{-0.3}  &  0.8^{+0.08}_{-0.1}  \\
2434  &  64.97  &  5.19  &  130^{+10}_{-6} &  0.007^{+0.002}_{-0.002}    &  0.035^{+0.004}_{-0.003}    &  0.5^{+0.1}_{-0.2}    &  2^{+0.1}_{-0.3}     &  1.1^{+0.2}_{-0.2}  &  0.8^{+0.1}_{-0.09}  \\
2435  &  66.09  &  5.24  &  130^{+6}_{-8}  &  0.01^{+0.004}_{-0.002}     &  0.037^{+0.003}_{-0.003}    &  1.7^{+0.2}_{-0.1}    &  1.4^{+0.3}_{-0.07}  &  0.8^{+0.3}_{-0.1}  &  0.8^{+0.1}_{-0.09}  \\
2436  &  66.48  &  5.34  &  130^{+7}_{-10} &  0.01^{+0.004}_{-0.002}     &  0.037^{+0.004}_{-0.004}    &  1.7^{+0.2}_{-0.1}    &  1.4^{+0.3}_{-0.09}  &  0.8^{+0.3}_{-0.1}  &  0.8^{+0.2}_{-0.1}   \\
2437  &  66.29  &  5.40  &  110^{+6}_{-20} &  0.008^{+0.004}_{-0.003}    &  0.031^{+0.004}_{-0.005}    &  1.7^{+0.3}_{-0.1}    &  1.7^{+0.2}_{-0.1}   &  0.8^{+0.3}_{-0.2}  &  0.8^{+0.2}_{-0.09}  \\
2438  &  66.25  &  5.35  &  130^{+6}_{-9}  &  0.006^{+0.001}_{-0.002}    &  0.034^{+0.003}_{-0.003}    &  1.1^{+0.1}_{-0.2}    &  1.7^{+0.2}_{-0.1}   &  1.1^{+0.1}_{-0.2}  &  0.8^{+0.1}_{-0.08}  \\
2439  &  66.79  &  5.38  &  130^{+10}_{-5} &  0.006^{+0.002}_{-0.002}    &  0.034^{+0.003}_{-0.003}    &  0.8^{+0.09}_{-0.2}   &  1.7^{+0.3}_{-0.1}   &  1.1^{+0.2}_{-0.2}  &  0.8^{+0.09}_{-0.09} \\
2440  &  67.54  &  5.45  &  130^{+7}_{-20} &  0.006^{+0.002}_{-0.003}    &  0.033^{+0.004}_{-0.005}    &  1.1^{+0.2}_{-0.2}    &  1.4^{+0.3}_{-0.09}  &  1.1^{+0.2}_{-0.3}  &  0.8^{+0.2}_{-0.1}   \\
2441  &  68.22  &  5.47  &  110^{+10}_{-4} &  0.005^{+0.002}_{-0.002}    &  0.028^{+0.003}_{-0.002}    &  \leqslant 0.34       &  2^{+0.2}_{-0.1}     &  1.1^{+0.2}_{-0.2}  &  0.8^{+0.1}_{-0.07}  \\
2442  &  68.70  &  5.48  &  90^{+20}_{-5}  &  0.007^{+0.003}_{-0.001}    &  0.028^{+0.006}_{-0.004}    &  1.1^{+0.2}_{-0.2}    &  1.7^{+0.2}_{-0.1}   &  0.8^{+0.3}_{-0.1}  &  1.1^{+0.07}_{-0.3}  \\
2443  &  69.02  &  5.53  &  110^{+8}_{-9}  &  0.008^{+0.004}_{-0.002}    &  0.03^{+0.004}_{-0.003}     &  \geqslant 1.9        &  2^{+0.3}_{-0.1}     &  0.8^{+0.3}_{-0.1}  &  0.8^{+0.1}_{-0.06}  \\
2444  &  69.31  &  5.57  &  130^{+6}_{-20} &  0.006^{+0.002}_{-0.002}    &  0.033^{+0.005}_{-0.005}    &  1.7^{+0.2}_{-0.2}    &  1.7^{+0.2}_{-0.3}   &  1.1^{+0.3}_{-0.2}  &  0.8^{+0.3}_{-0.09}  \\
2445  &  69.39  &  5.54  &  110^{+10}_{-7} &  0.008^{+0.004}_{-0.001}    &  0.03^{+0.005}_{-0.002}     &  \geqslant 1.9        &  \geqslant 2.1       &  0.8^{+0.3}_{-0.08} &  0.8^{+0.2}_{-0.07}  \\
2446  &  68.95  &  5.52  &  110^{+20}_{-8} &  0.005^{+0.002}_{-0.002}    &  0.028^{+0.006}_{-0.003}    &  0.8^{+0.3}_{-0.1}    &  \geqslant 2.1       &  1.1^{+0.2}_{-0.3}  &  0.8^{+0.3}_{-0.09}  \\
2447  &  69.39  &  5.48  &  130^{+6}_{-10} &  0.0041^{+0.0005}_{-0.002}  &  0.031^{+0.003}_{-0.004}    &  0.8^{+0.1}_{-0.1}    &  1.7^{+0.2}_{-0.2}   &  1.4^{+0.08}_{-0.3} &  0.8^{+0.2}_{-0.08}  \\
2448  &  69.57  &  5.51  &  130^{+7}_{-5}  &  0.004^{+0.0006}_{-0.002}   &  0.031^{+0.002}_{-0.002}    &  0.50^{+0.07}_{-0.1}  &  2^{+0.1}_{-0.2}     &  1.4^{+0.1}_{-0.3}  &  0.8^{+0.09}_{-0.06} \\
2449  &  69.31  &  5.49  &  130^{+10}_{-5} &  0.0063^{+0.003}_{-0.0006}  &  0.033^{+0.004}_{-0.002}    &  1.4^{+0.08}_{-0.1}   &  2^{+0.1}_{-0.2}     &  1.1^{+0.3}_{-0.06} &  0.8^{+0.1}_{-0.07}  \\
2450  &  69.96  &  5.48  &  130^{+4}_{-6}  &  0.0063^{+0.002}_{-0.0007}  &  0.033^{+0.002}_{-0.002}    &  1.4^{+0.07}_{-0.09}  &  1.7^{+0.1}_{-0.1}   &  1.1^{+0.2}_{-0.06} &  0.8^{+0.08}_{-0.05} \\
2451  &  70.68  &  5.38  &  110^{+20}_{-5} &  0.005^{+0.002}_{-0.002}    &  0.029^{+0.005}_{-0.002}    &  0.8^{+0.2}_{-0.2}    &  \geqslant 2         &  1.1^{+0.3}_{-0.3}  &  0.8^{+0.1}_{-0.09}  \\
2452  &  71.08  &  5.36  &  110^{+10}_{-5} &  0.005^{+0.002}_{-0.002}    &  0.029^{+0.004}_{-0.002}    &  0.8^{+0.2}_{-0.1}    &  2^{+0.2}_{-0.1}     &  1.1^{+0.2}_{-0.2}  &  0.8^{+0.1}_{-0.08}  \\
2453  &  71.09  &  5.38  &  90^{+9}_{-5}   &  0.0045^{+0.0009}_{-0.002}  &  0.023^{+0.003}_{-0.002}    &  0.5^{+0.2}_{-0.1}    &  \geqslant 2.2       &  1.1^{+0.1}_{-0.2}  &  0.8^{+0.2}_{-0.07}  \\
2454  &  70.97  &  5.36  &  90^{+20}_{-5}  &  0.004^{+0.002}_{-0.001}    &  0.026^{+0.006}_{-0.004}    &  0.8^{+0.2}_{-0.2}    &  1.7^{+0.3}_{-0.08}  &  1.1^{+0.3}_{-0.2}  &  1.1^{+0.07}_{-0.3}  \\
2455  &  70.59  &  5.40  &  110^{+5}_{-8}  &  0.005^{+0.002}_{-0.001}    &  0.028^{+0.003}_{-0.002}    &  1.7^{+0.2}_{-0.1}    &  2^{+0.1}_{-0.2}     &  1.1^{+0.3}_{-0.1}  &  0.8^{+0.1}_{-0.06}  \\
2456  &  70.25  &  5.31  &  110^{+6}_{-10} &  0.006^{+0.002}_{-0.001}    &  0.029^{+0.004}_{-0.004}    &  \geqslant 1.8        &  2^{+0.2}_{-0.2}     &  1.1^{+0.3}_{-0.2}  &  0.8^{+0.2}_{-0.08}  \\
2457  &  70.09  &  5.26  &  90^{+20}_{-4}  &  0.003^{+0.001}_{-0.001}    &  0.025^{+0.005}_{-0.003}    &  \leqslant 0.35       &  2^{+0.2}_{-0.2}     &  1.4^{+0.2}_{-0.3}  &  1.1^{+0.09}_{-0.3}  \\
2458  &  69.75  &  5.20  &  110^{+9}_{-10} &  0.004^{+0.001}_{-0.001}    &  0.027^{+0.004}_{-0.003}    &  1.1^{+0.2}_{-0.1}    &  \geqslant 2.1       &  1.4^{+0.3}_{-0.3}  &  0.8^{+0.2}_{-0.07}  \\
2459  &  69.38  &  5.16  &  110^{+9}_{-10} &  0.004^{+0.002}_{-0.001}    &  0.028^{+0.003}_{-0.003}    &  1.1^{+0.2}_{-0.1}    &  2^{+0.3}_{-0.1}     &  1.4^{+0.3}_{-0.2}  &  0.8^{+0.2}_{-0.07}  \\
2460  &  69.35  &  5.18  &  110^{+6}_{-10} &  0.0023^{+0.0003}_{-0.001}  &  0.026^{+0.002}_{-0.003}    &  \leqslant 0.37       &  2^{+0.2}_{-0.2}     &  1.7^{+0.09}_{-0.3} &  0.8^{+0.2}_{-0.07}  \\
2461  &  69.19  &  5.17  &  110^{+6}_{-8}  &  0.0036^{+0.001}_{-0.0007}  &  0.028^{+0.003}_{-0.002}    &  1.4^{+0.1}_{-0.2}    &  2^{+0.2}_{-0.2}     &  1.4^{+0.2}_{-0.1}  &  0.8^{+0.1}_{-0.08}  \\
2462  &  68.89  &  5.22  &  110^{+5}_{-4}  &  0.0023^{+0.0003}_{-0.001}  &  0.025^{+0.001}_{-0.002}    &  \leqslant 0.28       &  2^{+0.1}_{-0.1}     &  1.7^{+0.07}_{-0.3} &  0.8^{+0.06}_{-0.06} \\
2463  &  68.43  &  5.33  &  90^{+10}_{-5}  &  0.0029^{+0.001}_{-0.001}   &  0.022^{+0.004}_{-0.002}    &  \leqslant 0.41       &  \geqslant 2.1       &  1.4^{+0.2}_{-0.3}  &  0.8^{+0.2}_{-0.1}   \\
2464  &  68.01  &  5.29  &  110^{+5}_{-10} &  0.0036^{+0.001}_{-0.001}   &  0.027^{+0.003}_{-0.004}    &  1.4^{+0.1}_{-0.2}    &  2^{+0.2}_{-0.2}     &  1.4^{+0.2}_{-0.2}  &  0.8^{+0.2}_{-0.08}  \\
2465  &  67.03  &  5.34  &  110^{+6}_{-10} &  0.004^{+0.002}_{-0.001}    &  0.027^{+0.003}_{-0.004}    &  1.1^{+0.2}_{-0.1}    &  1.7^{+0.3}_{-0.08}  &  1.4^{+0.3}_{-0.2}  &  0.8^{+0.2}_{-0.1}   \\
2466  &  65.73  &  5.35  &  110^{+10}_{-5} &  0.0035^{+0.0009}_{-0.001}  &  0.027^{+0.003}_{-0.002}    &  0.8^{+0.1}_{-0.1}    &  \geqslant 2.1       &  1.4^{+0.2}_{-0.3}  &  0.8^{+0.1}_{-0.07}  \\
2467  &  64.25  &  5.34  &  110^{+3}_{-4}  &  0.0055^{+0.002}_{-0.0004}  &  0.029^{+0.002}_{-0.001}    &  \geqslant 1.9        &  2^{+0.1}_{-0.2}     &  1.1^{+0.2}_{-0.04} &  0.8^{+0.06}_{-0.05} \\
2468  &  63.14  &  5.30  &  110^{+6}_{-10} &  0.004^{+0.001}_{-0.001}    &  0.027^{+0.003}_{-0.004}    &  1.1^{+0.2}_{-0.1}    &  1.7^{+0.2}_{-0.09}  &  1.4^{+0.3}_{-0.2}  &  0.8^{+0.2}_{-0.09}  \\
2469  &  62.32  &  5.20  &  110^{+9}_{-5}  &  0.0036^{+0.0008}_{-0.001}  &  0.027^{+0.003}_{-0.003}    &  0.8^{+0.1}_{-0.1}    &  2^{+0.2}_{-0.2}     &  1.4^{+0.1}_{-0.3}  &  0.8^{+0.09}_{-0.1}  \\
2470  &  62.20  &  5.29  &  110^{+20}_{-6} &  0.004^{+0.001}_{-0.001}    &  0.027^{+0.005}_{-0.003}    &  \leqslant 0.41       &  2^{+0.3}_{-0.2}     &  1.4^{+0.3}_{-0.3}  &  0.8^{+0.1}_{-0.1}   \\
2471  &  62.36  &  5.40  &  110^{+20}_{-6} &  0.003^{+0.001}_{-0.001}    &  0.026^{+0.005}_{-0.004}    &  \leqslant 0.38       &  2^{+0.2}_{-0.3}     &  1.4^{+0.2}_{-0.3}  &  0.8^{+0.09}_{-0.3}  \\
2474  &  59.85  &  5.87  &  110^{+9}_{-3}  &  0.0078^{+0.003}_{-0.0006}  &  0.028^{+0.003}_{-0.002}    &  1.7^{+0.2}_{-0.1}    &  1.7^{+0.1}_{-0.1}   &  0.8^{+0.2}_{-0.05} &  0.8^{+0.05}_{-0.1}  \\
2475  &  58.91  &  6.04  &  110^{+20}_{-5} &  0.0076^{+0.004}_{-0.001}   &  0.027^{+0.005}_{-0.002}    &  1.7^{+0.2}_{-0.2}    &  2^{+0.2}_{-0.2}     &  0.8^{+0.3}_{-0.08} &  0.8^{+0.2}_{-0.2}   \\
2476  &  57.80  &  6.17  &  130^{+7}_{-20} &  0.006^{+0.002}_{-0.002}    &  0.029^{+0.004}_{-0.004}    &  1.4^{+0.2}_{-0.1}    &  1.4^{+0.2}_{-0.2}   &  1.1^{+0.3}_{-0.2}  &  0.8^{+0.2}_{-0.1}   \\
2477  &  56.43  &  6.18  &  110^{+7}_{-4}  &  0.0047^{+0.0008}_{-0.002}  &  0.025^{+0.002}_{-0.002}    &  0.8^{+0.1}_{-0.1}    &  2^{+0.2}_{-0.1}     &  1.1^{+0.1}_{-0.2}  &  0.8^{+0.2}_{-0.07}  \\
   \enddata
   \tablenotetext{a}{Bartels rotation number.}
   \tablenotetext{b}{Tilt angle, in units of degrees.}
   \tablenotetext{c}{HMF intensity at Earth, in units of nT.}
   \tablenotetext{d}{Normalization of the parallel diffusion coefficient, in units of \diffcoeff.}
   \tablenotetext{e}{Perpendicular mean free path at 1 GV and 5 GV at Earth, in units of AU. The uncertainty is computed by propagating the uncertainties on \cpa, \aperp\ and \bperp.}
   \tablenotetext{f}{Low-rigidity slope of the parallel diffusion coefficient.}
   \tablenotetext{g}{High-rigidity slope of the parallel diffusion coefficient.}
   \tablenotetext{h}{Low-rigidity slope of the perpendicular diffusion coefficient.}
   \tablenotetext{i}{High-rigidity slope of the perpendicular diffusion coefficient.}
   \tablecomments{A $\leqslant$ ($\geqslant$) symbol means that the best-fit value for the parameter coincides with the lower (upper) edge of the grid, so a lower (upper) limit is reported, corresponding to $\widehat{q}_{n} + q_{n,r}$ ($\widehat{q}_{n} - q_{n,l}$).}
\end{deluxetable*}

\startlongtable
\begin{deluxetable*}{c|cc|CCCCCCC}
   \tablecaption{
      Best-fit parameters used as input for numerical models with positive polarity.
      See Table \ref{tab:best-fit-pars:neg} for the description of the columns.
      \label{tab:best-fit-pars:pos}
   }
   \tablehead{
      \colhead{BR} & \colhead{\tilt} & \colhead{\cmf} & \colhead{\cpa} & \colhead{$\lambda_{\perp}$(1 GV)} & \colhead{$\lambda_{\perp}$(5 GV)} & \colhead{\apar} & \colhead{\bpar} & \colhead{\aperp} & \colhead{\bperp}
   }
   \startdata
2446  &  68.95  &  5.52  &  110^{+10}_{-9}  &  0.005^{+0.001}_{-0.001}     &  0.028^{+0.004}_{-0.003}    &  1.7^{+0.3}_{-0.3}   &  \geqslant 2.1        &  1.1^{+0.2}_{-0.1}     &  0.8^{+0.1}_{-0.06}    \\
2447  &  69.39  &  5.48  &  130^{+5}_{-10}  &  0.0041^{+0.0005}_{-0.001}   &  0.031^{+0.002}_{-0.003}    &  1.1^{+0.3}_{-0.2}   &  1.4^{+0.3}_{-0.2}    &  1.4^{+0.08}_{-0.2}    &  0.8^{+0.1}_{-0.08}    \\
2448  &  69.57  &  5.51  &  130^{+6}_{-6}   &  0.004^{+0.0006}_{-0.0006}   &  0.031^{+0.002}_{-0.002}    &  0.5^{+0.2}_{-0.1}   &  2^{+0.2}_{-0.2}      &  1.4^{+0.1}_{-0.1}     &  0.8^{+0.09}_{-0.04}   \\
2449  &  69.31  &  5.49  &  130^{+7}_{-7}   &  0.0041^{+0.0005}_{-0.001}   &  0.031^{+0.002}_{-0.003}    &  \leqslant 0.38      &  2^{+0.3}_{-0.2}      &  1.4^{+0.08}_{-0.2}    &  0.8^{+0.1}_{-0.05}    \\
2450  &  69.96  &  5.48  &  130^{+6}_{-9}   &  0.0041^{+0.0005}_{-0.001}   &  0.031^{+0.002}_{-0.003}    &  \leqslant 0.44      &  1.4^{+0.3}_{-0.3}    &  1.4^{+0.08}_{-0.2}    &  0.8^{+0.1}_{-0.09}    \\
2451  &  70.68  &  5.38  &  110^{+5}_{-5}   &  0.0055^{+0.0007}_{-0.0006}  &  0.029^{+0.002}_{-0.001}    &  1.4^{+0.3}_{-0.2}   &  \geqslant 2          &  1.1^{+0.08}_{-0.07}   &  0.8^{+0.07}_{-0.04}   \\
2452  &  71.08  &  5.36  &  110^{+5}_{-6}   &  0.0055^{+0.001}_{-0.0008}   &  0.029^{+0.002}_{-0.002}    &  1.7^{+0.3}_{-0.3}   &  \geqslant 2          &  1.1^{+0.1}_{-0.09}    &  0.8^{+0.09}_{-0.05}   \\
2453  &  71.09  &  5.38  &  90^{+7}_{-20}   &  0.004^{+0.001}_{-0.002}     &  0.023^{+0.003}_{-0.005}    &  1.7^{+0.3}_{-0.3}   &  \geqslant 2          &  1.1^{+0.2}_{-0.3}     &  0.8^{+0.3}_{-0.08}    \\
2454  &  70.97  &  5.36  &  90^{+20}_{-10}  &  0.004^{+0.002}_{-0.002}     &  0.023^{+0.006}_{-0.004}    &  1.7^{+0.3}_{-0.3}   &  \geqslant 2.1        &  1.1^{+0.3}_{-0.2}     &  0.8^{+0.2}_{-0.1}     \\
2455  &  70.59  &  5.40  &  110^{+5}_{-9}   &  0.0035^{+0.0007}_{-0.0007}  &  0.026^{+0.002}_{-0.002}    &  \geqslant 1.7       &  1.7^{+0.3}_{-0.2}    &  1.4^{+0.1}_{-0.1}     &  0.8^{+0.1}_{-0.07}    \\
2456  &  70.25  &  5.31  &  110^{+9}_{-20}  &  0.004^{+0.001}_{-0.001}     &  0.027^{+0.004}_{-0.005}    &  \geqslant 1.7       &  1.7^{+0.3}_{-0.2}    &  1.4^{+0.3}_{-0.2}     &  0.8^{+0.3}_{-0.1}     \\
2457  &  70.09  &  5.26  &  90^{+10}_{-5}   &  0.0029^{+0.0009}_{-0.0004}  &  0.025^{+0.004}_{-0.002}    &  0.5^{+0.3}_{-0.2}   &  1.7^{+0.2}_{-0.2}    &  1.4^{+0.2}_{-0.08}    &  1.1^{+0.08}_{-0.2}    \\
2458  &  69.75  &  5.20  &  110^{+10}_{-20} &  0.0023^{+0.0005}_{-0.0009}  &  0.026^{+0.003}_{-0.004}    &  \leqslant 0.44      &  \geqslant 2.1        &  1.7^{+0.1}_{-0.2}     &  0.8^{+0.2}_{-0.07}    \\
2459  &  69.38  &  5.16  &  110^{+8}_{-10}  &  0.0023^{+0.0004}_{-0.0007}  &  0.026^{+0.003}_{-0.003}    &  \leqslant 0.46      &  \geqslant 2.1        &  1.7^{+0.1}_{-0.2}     &  0.8^{+0.2}_{-0.06}    \\
2460  &  69.35  &  5.18  &  110^{+7}_{-10}  &  0.0023^{+0.0004}_{-0.0006}  &  0.026^{+0.002}_{-0.003}    &  0.5^{+0.3}_{-0.2}   &  \geqslant 2          &  1.7^{+0.1}_{-0.2}     &  0.8^{+0.2}_{-0.05}    \\
2461  &  69.19  &  5.17  &  110^{+5}_{-6}   &  0.0023^{+0.0003}_{-0.0004}  &  0.026^{+0.002}_{-0.002}    &  0.8^{+0.3}_{-0.3}   &  2^{+0.3}_{-0.3}      &  1.7^{+0.08}_{-0.1}    &  0.8^{+0.1}_{-0.05}    \\
2462  &  68.89  &  5.22  &  110^{+6}_{-5}   &  0.0023^{+0.0003}_{-0.0003}  &  0.025^{+0.002}_{-0.001}    &  \leqslant 0.47      &  2^{+0.2}_{-0.3}      &  1.7^{+0.09}_{-0.07}   &  0.8^{+0.07}_{-0.05}   \\
2463  &  68.43  &  5.33  &  90^{+10}_{-10}  &  0.0029^{+0.001}_{-0.0007}   &  0.022^{+0.004}_{-0.003}    &  0.8^{+0.3}_{-0.3}   &  \geqslant 2.1        &  1.4^{+0.2}_{-0.1}     &  0.8^{+0.1}_{-0.09}    \\
2464  &  68.01  &  5.29  &  110^{+7}_{-10}  &  0.0023^{+0.0004}_{-0.0006}  &  0.025^{+0.002}_{-0.003}    &  1.7^{+0.3}_{-0.3}   &  2^{+0.3}_{-0.3}      &  1.7^{+0.1}_{-0.2}     &  0.8^{+0.2}_{-0.06}    \\
2465  &  67.03  &  5.34  &  110^{+6}_{-9}   &  0.0023^{+0.0004}_{-0.0005}  &  0.025^{+0.002}_{-0.002}    &  0.5^{+0.2}_{-0.3}   &  2^{+0.2}_{-0.3}      &  1.7^{+0.1}_{-0.1}     &  0.8^{+0.1}_{-0.05}    \\
2466  &  65.73  &  5.35  &  130^{+6}_{-10}  &  0.0027^{+0.0004}_{-0.0007}  &  0.029^{+0.002}_{-0.003}    &  1.7^{+0.3}_{-0.3}   &  1.4^{+0.2}_{-0.2}    &  1.7^{+0.1}_{-0.2}     &  0.8^{+0.1}_{-0.09}    \\
2467  &  64.25  &  5.34  &  110^{+10}_{-10} &  0.0023^{+0.0006}_{-0.0006}  &  0.025^{+0.003}_{-0.003}    &  \leqslant 0.4       &  \geqslant 2          &  1.7^{+0.2}_{-0.2}     &  0.8^{+0.2}_{-0.2}     \\
2468  &  63.14  &  5.30  &  110^{+6}_{-9}   &  0.0023^{+0.0003}_{-0.0005}  &  0.025^{+0.002}_{-0.002}    &  \leqslant 0.48      &  2^{+0.3}_{-0.3}      &  1.7^{+0.09}_{-0.1}    &  0.8^{+0.1}_{-0.05}    \\
2469  &  62.32  &  5.20  &  130^{+20}_{-20} &  0.003^{+0.001}_{-0.001}     &  0.03^{+0.005}_{-0.005}     &  1.7^{+0.3}_{-0.3}   &  1.4^{+0.3}_{-0.2}    &  1.7^{+0.2}_{-0.2}     &  0.8^{+0.2}_{-0.2}     \\
2470  &  62.20  &  5.29  &  130^{+8}_{-10}  &  0.0027^{+0.0003}_{-0.0008}  &  0.03^{+0.002}_{-0.003}     &  \leqslant 0.39      &  1.1^{+0.2}_{-0.04}   &  1.7^{+0.06}_{-0.2}    &  0.8^{+0.07}_{-0.1}    \\
2471  &  62.36  &  5.40  &  110^{+20}_{-7}  &  0.0035^{+0.002}_{-0.0006}   &  0.026^{+0.005}_{-0.003}    &  \leqslant 0.49      &  2^{+0.3}_{-0.3}      &  1.4^{+0.3}_{-0.1}     &  0.8^{+0.1}_{-0.2}     \\
2474  &  59.85  &  5.87  &  110^{+5}_{-3}   &  0.005^{+0.0005}_{-0.0003}   &  0.026^{+0.001}_{-0.001}    &  1.7^{+0.3}_{-0.3}   &  1.7^{+0.2}_{-0.3}    &  1.1^{+0.07}_{-0.04}   &  0.8^{+0.04}_{-0.07}   \\
2475  &  58.91  &  6.04  &  110^{+20}_{-5}  &  0.0049^{+0.002}_{-0.0006}   &  0.025^{+0.005}_{-0.002}    &  1.4^{+0.3}_{-0.3}   &  \geqslant 2          &  1.1^{+0.3}_{-0.08}    &  0.8^{+0.08}_{-0.1}    \\
2476  &  57.80  &  6.17  &  130^{+4}_{-9}   &  0.0036^{+0.0003}_{-0.0008}  &  0.027^{+0.001}_{-0.002}    &  0.5^{+0.3}_{-0.2}   &  1.1^{+0.2}_{-0.05}   &  1.4^{+0.06}_{-0.1}    &  0.8^{+0.07}_{-0.06}   \\
2477  &  56.43  &  6.18  &  110^{+20}_{-20} &  0.005^{+0.002}_{-0.001}     &  0.025^{+0.005}_{-0.004}    &  1.7^{+0.3}_{-0.3}   &  \geqslant 2          &  1.1^{+0.3}_{-0.1}     &  0.8^{+0.2}_{-0.1}     \\
2478  &  55.22  &  6.29  &  110^{+6}_{-9}   &  0.003^{+0.0004}_{-0.0006}   &  0.023^{+0.002}_{-0.002}    &  \leqslant 0.51      &  \geqslant 2          &  1.4^{+0.08}_{-0.1}    &  0.8^{+0.1}_{-0.04}    \\
2479  &  54.36  &  6.33  &  130^{+10}_{-20} &  0.0023^{+0.0005}_{-0.001}   &  0.025^{+0.003}_{-0.004}    &  \geqslant 1.7       &  1.7^{+0.3}_{-0.3}    &  1.7^{+0.1}_{-0.3}     &  0.8^{+0.3}_{-0.1}     \\
2480  &  54.19  &  6.36  &  150^{+10}_{-20} &  0.0026^{+0.0004}_{-0.001}   &  0.028^{+0.003}_{-0.004}    &  \leqslant 0.44      &  1.4^{+0.3}_{-0.3}    &  1.7^{+0.09}_{-0.2}    &  0.8^{+0.1}_{-0.1}     \\
2481  &  54.24  &  6.43  &  130^{+6}_{-5}   &  0.0035^{+0.0005}_{-0.0004}  &  0.026^{+0.002}_{-0.001}    &  \leqslant 0.4       &  2^{+0.2}_{-0.3}      &  1.4^{+0.09}_{-0.08}   &  0.8^{+0.08}_{-0.07}   \\
2482  &  54.08  &  6.51  &  150^{+6}_{-6}   &  0.0039^{+0.0006}_{-0.0005}  &  0.03^{+0.002}_{-0.001}     &  0.8^{+0.2}_{-0.2}   &  1.4^{+0.2}_{-0.3}    &  1.4^{+0.09}_{-0.08}   &  0.8^{+0.06}_{-0.05}   \\
2483  &  53.42  &  6.64  &  170^{+9}_{-20}  &  0.0044^{+0.0009}_{-0.001}   &  0.033^{+0.002}_{-0.004}    &  1.7^{+0.3}_{-0.3}   &  \leqslant 0.46       &  1.4^{+0.1}_{-0.2}     &  0.8^{+0.07}_{-0.1}    \\
2484  &  52.55  &  6.72  &  150^{+20}_{-5}  &  0.006^{+0.002}_{-0.0006}    &  0.031^{+0.004}_{-0.002}    &  \geqslant 1.8       &  1.4^{+0.3}_{-0.3}    &  1.1^{+0.2}_{-0.06}    &  0.8^{+0.08}_{-0.1}    \\
2485  &  51.26  &  6.67  &  150^{+10}_{-4}  &  0.006^{+0.002}_{-0.0005}    &  0.031^{+0.003}_{-0.001}    &  \geqslant 1.8       &  1.4^{+0.2}_{-0.3}    &  1.1^{+0.2}_{-0.06}    &  0.8^{+0.06}_{-0.09}   \\
2486  &  50.12  &  6.66  &  150^{+8}_{-5}   &  0.006^{+0.001}_{-0.0006}    &  0.031^{+0.002}_{-0.001}    &  \geqslant 1.7       &  1.4^{+0.3}_{-0.3}    &  1.1^{+0.1}_{-0.07}    &  0.8^{+0.06}_{-0.06}   \\
2487  &  49.89  &  6.60  &  170^{+9}_{-10}  &  0.0044^{+0.0005}_{-0.002}   &  0.033^{+0.002}_{-0.004}    &  \leqslant 0.36      &  \leqslant 0.5        &  1.4^{+0.07}_{-0.3}    &  0.8^{+0.07}_{-0.1}    \\
2488  &  49.67  &  6.56  &  170^{+8}_{-7}   &  0.007^{+0.001}_{-0.001}     &  0.036^{+0.002}_{-0.002}    &  1.7^{+0.2}_{-0.2}   &  \leqslant 0.48       &  1.1^{+0.1}_{-0.1}     &  0.8^{+0.05}_{-0.1}    \\
2489  &  49.34  &  6.52  &  190^{+9}_{-10}  &  0.005^{+0.0005}_{-0.002}    &  0.038^{+0.002}_{-0.004}    &  \leqslant 0.35      &  \leqslant 0.46       &  1.4^{+0.06}_{-0.3}    &  0.8^{+0.07}_{-0.1}    \\
2490  &  49.14  &  6.50  &  190^{+10}_{-8}  &  0.008^{+0.002}_{-0.001}     &  0.041^{+0.004}_{-0.004}    &  1.7^{+0.2}_{-0.2}   &  \leqslant 0.5        &  1.1^{+0.2}_{-0.1}     &  0.8^{+0.06}_{-0.2}    \\
2491  &  48.94  &  6.47  &  190^{+20}_{-10} &  0.008^{+0.003}_{-0.002}     &  0.041^{+0.005}_{-0.006}    &  1.4^{+0.3}_{-0.2}   &  \leqslant 0.46       &  1.1^{+0.3}_{-0.2}     &  0.8^{+0.06}_{-0.3}    \\
2492  &  48.80  &  6.44  &  210^{+9}_{-20}  &  0.006^{+0.001}_{-0.002}     &  0.038^{+0.005}_{-0.004}    &  0.5^{+0.2}_{-0.2}   &  \geqslant 2          &  1.4^{+0.1}_{-0.3}     &  0.50^{+0.3}_{-0.05}   \\
2493  &  48.59  &  6.47  &  190^{+20}_{-10} &  0.008^{+0.003}_{-0.003}     &  0.041^{+0.005}_{-0.006}    &  1.1^{+0.3}_{-0.2}   &  \leqslant 0.48       &  1.1^{+0.3}_{-0.2}     &  0.8^{+0.06}_{-0.3}    \\
2494  &  47.84  &  6.40  &  210^{+10}_{-10} &  0.009^{+0.003}_{-0.002}     &  0.041^{+0.005}_{-0.003}    &  1.7^{+0.2}_{-0.3}   &  \geqslant 2.1        &  1.1^{+0.2}_{-0.1}     &  0.50^{+0.2}_{-0.04}   \\
2495  &  47.01  &  6.37  &  210^{+9}_{-20}  &  0.0056^{+0.0007}_{-0.002}   &  0.038^{+0.005}_{-0.004}    &  \leqslant 0.36      &  \geqslant 2.1        &  1.4^{+0.08}_{-0.3}    &  0.50^{+0.3}_{-0.04}   \\
2496  &  46.45  &  6.33  &  210^{+9}_{-20}  &  0.0056^{+0.0007}_{-0.002}   &  0.038^{+0.005}_{-0.004}    &  \leqslant 0.36      &  \geqslant 2.1        &  1.4^{+0.08}_{-0.3}    &  0.50^{+0.3}_{-0.04}   \\
2497  &  45.86  &  6.30  &  210^{+10}_{-20} &  0.009^{+0.002}_{-0.002}     &  0.041^{+0.005}_{-0.004}    &  1.7^{+0.2}_{-0.3}   &  \geqslant 2.1        &  1.1^{+0.2}_{-0.2}     &  0.50^{+0.2}_{-0.04}   \\
2498  &  45.47  &  6.20  &  210^{+10}_{-10} &  0.009^{+0.002}_{-0.003}     &  0.042^{+0.005}_{-0.004}    &  1.4^{+0.2}_{-0.2}   &  \geqslant 2.1        &  1.1^{+0.2}_{-0.2}     &  0.50^{+0.2}_{-0.04}   \\
2499  &  45.05  &  6.15  &  190^{+20}_{-8}  &  0.013^{+0.006}_{-0.002}     &  0.046^{+0.006}_{-0.006}    &  \geqslant 1.7       &  0.5^{+0.3}_{-0.3}    &  0.8^{+0.3}_{-0.09}    &  0.8^{+0.08}_{-0.3}    \\
2500  &  44.38  &  6.12  &  230^{+10}_{-20} &  0.01^{+0.003}_{-0.005}      &  0.047^{+0.006}_{-0.005}    &  1.4^{+0.2}_{-0.3}   &  2^{+0.3}_{-0.2}      &  1.1^{+0.2}_{-0.3}     &  0.50^{+0.2}_{-0.05}   \\
2501  &  43.52  &  6.04  &  210^{+20}_{-20} &  0.014^{+0.006}_{-0.003}     &  0.046^{+0.007}_{-0.004}    &  \geqslant 1.8       &  \geqslant 2          &  0.8^{+0.3}_{-0.1}     &  0.50^{+0.2}_{-0.05}   \\
2502  &  42.56  &  5.92  &  210^{+20}_{-20} &  0.015^{+0.006}_{-0.004}     &  0.047^{+0.007}_{-0.005}    &  \geqslant 1.7       &  \geqslant 2          &  0.8^{+0.3}_{-0.2}     &  0.50^{+0.3}_{-0.05}   \\
2503  &  41.26  &  5.80  &  210^{+10}_{-20} &  0.015^{+0.006}_{-0.003}     &  0.048^{+0.006}_{-0.004}    &  \geqslant 1.8       &  \geqslant 2.1        &  0.8^{+0.3}_{-0.1}     &  0.50^{+0.2}_{-0.04}   \\
2504  &  39.91  &  5.67  &  210^{+10}_{-20} &  0.01^{+0.003}_{-0.004}      &  0.046^{+0.006}_{-0.005}    &  0.5^{+0.2}_{-0.2}   &  \geqslant 2.1        &  1.1^{+0.2}_{-0.3}     &  0.50^{+0.2}_{-0.05}   \\
2505  &  38.61  &  5.59  &  210^{+10}_{-20} &  0.01^{+0.002}_{-0.004}      &  0.047^{+0.006}_{-0.005}    &  0.8^{+0.2}_{-0.2}   &  \geqslant 2.1        &  1.1^{+0.1}_{-0.3}     &  0.50^{+0.3}_{-0.05}   \\
2506  &  37.77  &  5.52  &  170^{+20}_{-20} &  0.013^{+0.005}_{-0.005}     &  0.046^{+0.006}_{-0.007}    &  1.1^{+0.3}_{-0.2}   &  0.5^{+0.3}_{-0.3}    &  0.8^{+0.3}_{-0.3}     &  0.8^{+0.1}_{-0.3}     \\
   \enddata
\end{deluxetable*}

\end{document}